\documentclass[useAMS,usenatbib]{mn2e}

\usepackage{epsf,rotating}
\usepackage{amsmath}
\usepackage{subfigure}
\usepackage[]{hyperref}
\usepackage{amssymb}
\usepackage{color}
\usepackage{verbatimbox}
\usepackage{bm}
\usepackage{bigfoot}

\newcommand{\msolar}{\;{\rm M}_{\odot}}

\newcommand{\ion}[2]{\hbox{#1\,{\sc #2}}}

\newcommand{\aap}{A\&A}

\newcommand{\apj}{ApJ}

\newcommand{\mnras}{MNRAS}

\newcommand{\fesc}{f_{\mathrm{esc}}}

\newcommand{\rion}{{R_{\rm ion}}}
\newcommand{\rrec}{{R_{\rm rec}}}
\newcommand{\aion}{{A_{\rm ion}}}
\newcommand{\cion}{{C_{\rm ion}}}
\title[EoR 21cm Forecasting]{Epoch of Reionisation 21cm Forecasting From MCMC-Constrained Semi-Numerical Models}

\author[S. Hassan et al.]{
\parbox[t]{\textwidth}{\vspace{-1cm}
Sultan Hassan$^1$, Romeel Dav\'e$^{1,2,3}$, Kristian Finlator$^4$, Mario G. Santos$^{1,5}$}
\\
\\$^1$ University of the Western Cape, Bellville, Cape Town, 7535, South Africa
\\$^2$ South African Astronomical Observatories, Observatory, Cape Town, 7925, South Africa
\\$^3$ African Institute for Mathematical Sciences, Muizenberg, Cape Town, 7945, South Africa
\\$^4$ New Mexico State University, Las Cruces, NM 88003, United States
\\$^5$ SKA SA, The Park, Park Road, Pinelands 7405, South Africa
}

\begin{document}

\maketitle

 \begin{abstract}
The recent low value of \citet{planck16} integrated optical depth
to Thomson scattering suggests that the reionisation occurred fairly
suddenly, disfavoring extended reionisation scenarios. This will
have a significant impact on the 21cm power spectrum. Using a
semi-numerical framework, we improve our model from \citet{hassan16}
to include time-integrated ionisation and recombination effects, and
find that this leads to more sudden reionisation.  It also yields
larger \ion{H}{ii} bubbles which leads to an order of magnitude more 21cm
power on large scales, while suppressing the small scale ionisation power.  Local
fluctuations in the neutral hydrogen density play the dominant role
in boosting the 21cm power spectrum on large scales, while recombinations are subdominant.
We use a Monte Carlo Markov Chain approach to constrain our model
to observations of the star formation rate functions at $z=6,7,8$
from \citet{bouw15}, the \citet{planck16} optical depth measurements,
and the \citet{bec13} ionising emissivity data at $z\sim 5$. We
then use this constrained model to perform 21cm forecasting for
LOFAR, HERA, and SKA in order to determine how well such data
can characterise the sources driving reionisation.  We find that
the 21cm power spectrum alone can somewhat constrain the halo mass
dependence of ionising sources, the photon escape fraction
and ionising amplitude, but combining the 21cm data with other
current observations enables us to separately constrain all these
parameters.  Our framework illustrates how 21cm data can play a
key role in understanding the sources and topology of reionisation
as observations improve.
\end{abstract}


\section{Introduction}

The redshifted 21cm neutral hydrogen line from the Epoch of
Reionisation (EoR) provides numerous astrophysical and cosmological
information about the formation and evolution of the first stars
and galaxies~\citep{bar01}.  Many ongoing and forthcoming experiments
such as the Low Frequency Array (LOFAR)\footnote{http://www.lofar.org/},
the Hydrogen Epoch of Reionisation Array
(HERA)\footnote{http://reionisation.org}, and the Square Kilometer
Array (SKA-Low)\footnote{https://www.skatelescope.org}, are devoted
to observing the dense neutral hydrogen gas that traces the cosmic
web at redshifts beyond 7.  While current experiments only yield
upper limits to the measurements of the 21cm power spectrum, these
future experiments are likely to provide a detection in the near
future.  It is thus important to develop robust and comprehensive
theoretical models that can utilise such information, along with
observations from other wavelengths and facilities, in order to
optimally constrain the physical processes driving reionisation.

The recent low value of \citet{planck16} integrated optical depth
to Thomson scattering suggests that the EoR may have occurred more
suddenly, and at much later times, than what was previously
believed~\citep{hin13}. The low value of $\tau = 0.058 \pm 0.012$
prefers EoR models with late onset and shorter duration. This, in
turn, is expected to have a significant impact on the expected 21cm
signal and its evolution.  Proper modeling of the sources and sinks
of ionising photons during the EoR is required to accurately model
the \ion{H}{ii} bubbles and study their sizes and distributions.
Doing so will enable us to connect the observed 21cm power spectrum
with the physical properties of the sources and sinks of ionising
photons during the EoR.

There are several major challenges to modeling the EoR and its
redshifted 21cm signal, driven by the requirements for accurately
modeling the power spectrum of \ion{H}{i} on large scales.  These
requirements include: (i) large volumes ($\sim$ 500 Mpc) in order
to capture the large scale \ion{H}{i} fluctuations that will be
detected in upcoming 21cm observations; (ii) high resolution that
is sufficient to resolve the ionising sources and self-shielding
systems on sub-kpc scales~\citep{ian15}; and (iii) accurate tracking
of the ionising radiation and other feedback processes from the
sub-kpc up to Mpc scales.  For these reasons, self-consistently
simulating the EoR represents an immense computational challenge
that no current model has been able to fully meet.

Nonetheless, great progress has been made in simulating the EoR on
both small and large scales.  Hydrodynamic simulations that
self-consistently incorporate radiative transfer
(\citealt{gne00,gne14,pawlik08,fin09,fin13,katz16}) have sufficient
resolution to model the sources of reionisation direction, and to
propagate the emitted radiation through the intergalactic medium
(IGM) with minimal physical assumptions.  However, owing to
computational limitations they are currently restricted to volumes
smaller than $\sim 20$~Mpc in order to resolve all atomically-cooling
halos.  An alternative approach is to post-process simulated density
fields with radiative transfer (\citealt{ raz02,mel06,mcq07,
thom09,ian14,bau15}), which allows access to larger volumes but
does not self-consistently account for thermal, ionisation, and
chemical feedback effects on galaxy formation.  Semi-analytical EoR
models ~\citep{mit11,mit13} are very successful in studying and
constraining the globally averaged astrophysical quantities and
parameters during EoR ~\citep{mit15,mit15b} based on current
observations, but lack the dynamic range to study 21cm fluctuations.
Finally, on the very largest scales, semi-numerical models
(\citealt{mesinger07,zah07,cho09,san10}) based on quasi-linear
density evolution with coarse modeling of the source population are
able to access volumes sufficient to make 21cm predictions relevant
to upcomgin observations, but must employ simple parameterised
approximations for the source and sink populations.  Nonetheless,
such semi-numerical models, with appropriate tuning, can reproduce
similar reionisaion histories as obtained by full radiative transfer
simulations (\citealt{zah11,maj14}).

Semi-numerical models are most ideally suited for studying the
large-scale ($\ga 1~Mpc$) 21cm power spectrum that will be measured
with upcoming radio facilities, but they make many simplifying
assumptions.  In particular, they must assume parameterisations for
the relationship between halo mass and ionising luminosity, and the
relationship between the large-scale density field and the recombination
rate that emerges from small-scale clumping.  Also, current
semi-numerical codes treat a single cell as either fully neutral
or fully ionised, hence they must choose some condition to assign
that cell as ionised.  This third condition is an algorithmic choice,
but the first two connect to physics, as they provide an opportunity
to constrain astrophysical quantities associated with EoR sources
and sinks based on 21cm and other EoR observations.

In \citet{hassan16}, we focused on improving the physical
parameterisations of the source and sink populations in the
semi-numerical model {\sc SimFast21} by employing parameterised
results from high-resolution radiative hydrodynamic simulations.
This enabled greater physical realism of parameterised source and
sink populations compared to previous approaches that had used a
linear relationship between halo mass and luminosity, and did not
include recombinations.  To do this, we obtained parametrizations
for the ionisation rate $\rion$ and recombination rate $\rrec$ as
functions of halo mass, overdensity, and redshift, extracted from
high resolution radiative transfer hydrodynamic simulation~\citep{fin15}
(hearafter 6/256-RT) and larger-volume hydrodynamic
simulation~\citep{dav13} (hereafter 32/512).  We then implemented
these parametrizations into {\sc SimFast21}, and identified ionised
regions where the ionisation rate exceeded the recombination rate.
This more realistic modeling replaces the canonical efficiency
parameter $\zeta$ approach in previous semi-numerical EoR modeling.
In particular, we found that the $\rion$ scales super-linearly with
halo mass ($\rion \propto M^{1.4}_{h}$) in contrast to the typically
assumed linear relationship between the efficiency parameter $\zeta$
and halo mass.  We showed that using these new parametrizations
($\rion$ and $\rrec$) allows us to simultaneously match various EoR
key observables with a relatively low escape fraction, independent
of halo mass and redshift.  We also found that the $\rion$ boosts
the small scale 21cm power spectrum while $\rrec$ suppresses the
21cm power on large scales during cosmic reionisations.

\citet{hassan16} thus improved upon the first two major uncertainties
in semi-numerical models, namely the ionisation and recombinations.
However, this work still assumed an ionisation condition based on
the {\it instantaneous} balance between ionisations and recombinations
-- in other words, if there were instantaneously more ionisations
than recombinations, that volume of space was considered fully
ionised.  However, this is not physically fully accurate, because
the excess ionising photons in such regions still must ionise the
neutral hydrogen atoms in that region. The instantaneous criterion
thus does not account for partial ionisation of a given cell, thus
it underestimates the total number of photons required.  In the
limiting case where reionisation proceeds quickly, this may not be
a bad approximation, but ideally we aim to relax this instantaneous
assumption.  In essence, it is likely that our ionisations were too
efficient, which can affect the topology and duration of the EoR along
with our constraints on $\fesc$.

In this paper, we improve upon our previous ionisation condition
by tracking the actual number of neutral hydrogen atoms, ionising
photons, and recombinations.  This leads to a time dependent
ionisation condition that is analogous to the well-known ionisation
balance equation.  With this, it turns out that reionisation occurs
more suddenly, as preferred by the recent \citet{planck16} constraints,
but requires a higher escape fraction.  We compare this to our
previous Instantaneous EoR model \citet{hassan16} in terms of their
\ion{H}{ii} bubble sizes, EoR history, and 21cm power spectra.

Ideally, we would like to use the 21cm power spectra and other
observations to provide constraints on the nature of the source
population, in particular its relationship to the halo population.
In \citet{hassan16}, we manually constrained the relationship between
ionising emissivity and halo mass versus observations, since we
only had one free parameter, namely the escape fraction of ionising
photons.  This was because we had fixed the characteristics of the
source population based on our radiative hydrodynamic simulations.
Here we would like to relax this assumption, and determine how well
we can constrain the source population characteristics directly
from observations.  To do this, we consider a generalised model
with three free parameters: the escape photon fraction $\rm \fesc$,
the ionising emissivity amplitude $\rm A_{ion}$ ($\rion$ amplitude),
and the ionising emissivity-halo mass power-law index $\rm C_{ion}$.
We note that $\rm C_{ion}$ can represent the power-law mass dependence
of either the amplitude $\rm A_{ion}$ or the escape fraction $\fesc$;
in our current approach, these two quantities are degenerate.  To
constrain these parameters, we perform a Bayesian Monte Carlo Markov
Chain (MCMC) search against current EoR observations.  We then
forecast how these constraints will be improved by upcoming 21cm
observations from LOFAR, HERA, and SKA-Low.  By considering all
such observations, we determine how well we can constrain the EoR
source population as characterised by our three free parameters.

This paper is organized as follows: In section \ref{sec:sims}, we
introduce our previous Instantaneous EoR model and our new Time-integrated model.
We study and compare these models' impact on various EoR observables
including the 21cm power spectrum in section\ref{sec:impact}.  In
section \ref{sec:effects}, we create several EoR models to study
their effects on the 21cm power spectrum. In section\ref{sec:calibration},
we calibrate the Time-integrated model to various EoR observations. We perform
the 21cm forecasting in section\ref{sec:21cmforecating} and draw
our concluding remarks in section \ref{sim:conclusion}.

Throughout this work, we adopt a $\Lambda$CDM cosmology in which
$\Omega_{\rm M}=0.3$, $\Omega_{\rm \Lambda}=0.7$, $h\equiv H_0/(100
\, \mathrm{km/s/Mpc})=0.7$, a primordial power spectrum index
$n=0.96$, an amplitude of the mass fluctuations scaled to $\sigma_8=0.8$,
and $\Omega_b=0.045$. We quote all results in comoving units, unless
otherwise stated.

\section{Simulations}\label{sec:sims}

We use a semi-numerical code {\sc SimFast21}~(\citealt{san10}), which
we briefly review here.  {\sc SimFast21} simulation begins by
generating the density field from a Gaussian distribution using a
Monte-Carlo approach. The generated density field will then be
dynamically evolved from linear to non-linear regime by applying
the \citet{zeldovich70} approximation. The dark matter halos are
generated using the well-known excursion-set formalism. In the
standard {\sc SimFast21}, the ionised regions are identified using
the excursion-set formalism based on a
constant efficiency parameter $\zeta$. In the original {\sc SimFast21}
code, the ionisation condition
compares the amount of collapsed dark matter halo
$f_{coll}$ to the efficiency parameter $\zeta$ -- any region will be
flagged as ionised if:
\begin{equation}\label{eq:zeta}
 f_{coll} \geq  \zeta^{-1}.
\end{equation}
The efficiency parameter $\zeta$ is a model free parameter
which can be tuned to match some observations. This condition generates
the ionisation field, which may be used along with the density
field to obtain the 21cm brightness temperature. We refer the
reader to \citet{san10} for more details on this model and the code
algorithm. 

We now describe our two extensions to {\sc SimFast21}.  The first
was presented in \citet{hassan16}, which we review next, and
incorporates the ionisation and recombination rate parameterisations
taken from hydrodynamic simulations, but utilises an instantaneous
ionisation condition.  We then describe our further extension here
in order to improve the ionisation condition by tracking the neutral
fraction in a time-integrated manner.  We will call this the
``Instantaneous ionisation" and "Time-integrated ionisation" models.

\subsection{Instantaneous ionisation model}

As described in \citet{hassan16}, here we replace the efficiency
parameter $\zeta$  with direct parameterisations of the ionisation
rate $\rion$ and recombination rate $\rrec$ as functions of halo
mass $M_{h}$, overdensity $\Delta$, and redshift $z$, taken from our
6/256-RT and 32/512 simulations. Our best-fit non-linear ionisation rate
$\rion$ parametrization takes the following form:
\begin{equation}\label{eq:nion}
\mathrm{\frac{\rion}{M_{h}} =  A_{ion}\, (1 + z)^{D_{ion}} \, ( M_{h}/B_{ion} )^{C_{ion}} \, \exp\left( -( B_{ion}/M_{h})^{3.0} \right) } ,
\end{equation}
where $\mathrm{A_{ion}} =1.08\times 10^{40}  \msolar^{-1} $s$^{-1}$,
$\mathrm{B_{ion}} = 9.51\times 10^{7}\msolar$, $\mathrm{C_{ion}} =
0.41$ and $\mathrm{D_{ion}} = 2.28$. Meanwhile, we parameterise the recombination
rate as:
\begin{equation}\label{eq:rrec}
\mathrm{\frac{\rrec}{V} =  A_{rec}(1+ z)^{D_{rec}}  \left[\frac{\left( \Delta/B_{rec} \right)^{C_{rec}}}{1+ \left( \Delta/B_{rec} \right)^{C_{rec}} } \right]^{4} }\, , 
\end{equation}
where $\mathrm{A_{rec}} = 9.85 \times 10^{-24} $cm$^{-3}$s$^{-1}$,
$\mathrm{B_{rec}} = 1.76 $, $\mathrm{C_{rec}} = 0.82$,
$\mathrm{D_{rec}}=5.07$. 

Our ionisation condition is taken to be
\begin{equation}\label{eq:inst}\rm
\fesc R_\mathrm{ion,V} \geq R_\mathrm{rec,V}\, ,
\end{equation}
where 
\begin{equation*}\rm 
R_\mathrm{ion,V} = \int_{dn}\int_{V} \frac{dn}{dM_{h}} \rion(M_{h},z)\,dM_{h} \,dV\, , 
\end{equation*} 
and 
\begin{equation*}\rm
R_\mathrm{rec,V}=  \int_{V} \,\rrec(\Delta,z)\, dV\,. 
\end{equation*}
In above expressions, the $\rm \fesc$ is the photon escape fraction, V is the spherical region volume specified by the excursion set-formalism, and n is the number density of halos. With these volume integrals, the $\rm R_\mathrm{ion,V} $ represents the total ionisation rate from all sources and $\rm R_\mathrm{rec,V}$ is the maximum recombination rate in that volume V.  Cells in a given volume V satisfying this criterion (equation~\ref{eq:inst}) are considered fully ionised, otherwise they are fully neutral.

Using this model, it has been shown in ~\citet{hassan16} that one
can match simultaneously several EoR key observables, such as
~\citet{planck15} optical depth, ~\citet{bec13} ionising emissivity
and ~\citet{fan06} filling factor measurement, by only a constant
$\rm \fesc = 4-6\%$ independent of halo mass or redshift. We refer the
reader to ~\citet{hassan16} for more details about the model and
these new parametrizations.

\subsection{Time-integrated ionisation model}

The model in ~\citet{hassan16} (our Instantaneous EoR model) has several
drawbacks.  First, the ionisation condition, equation~\eqref{eq:inst},
is an instantaneous criterion which compares the escaped ionisation
rate $\fesc\rion$ with the recombination rate $\rrec$, instead of
comparing the actual numbers of ionising photons to that of
recombinations. Second, the ionisation condition, equation~\eqref{eq:inst},
assumes maximum recombination rate $\rrec$ from all cells/regions
as if they were fully ionised.

To improve on these, we modify the ionisation condition to account
for the evolving neutral hydrogen fraction, which allows us to
account for the existing number of hydrogen atoms in each region
as well as to compute the recombination based on the current ionised
fraction.  Hence we now employ a time dependent integral ionisation
condition:
\begin{equation}
\label{eq:new_con}\rm 
\fesc R_\mathrm{ion,V} \,\, dt \geq {I\!R}_\mathrm{rec,V}\,\, dt +  N_{HI,V}\, ,
\end{equation}
where
\begin{equation*}\rm
 {I\!R}_\mathrm{rec,V}=  \int_{V}  x_{HII} \,\rrec(\Delta,z)\, dV\, , 
\end{equation*}
and
\begin{equation*}\rm 
N_{HI,V} =  \int_{V}  (1-x_{HII})\, N_{H}\, dV. 
\end{equation*}
The $\rm x_{HII}$ and $\rm N_{H}$ here are the ionisation fraction and the total number of hydrogen in cells respectively. The dt represents the time duration between successive snapshots. We apply this new condition (equation~\ref{eq:new_con}) as follows: for each cell, we first compute the integrand of the LHS ($\rm \fesc \rion dt$) and RHS ($\rm x_{HII} \rrec dt+ (1-x_{HII}) NH$). We then apply the excursion set formalism to perform the spherical volume integral. Once again, the ionisation fraction is set to 1 (fully ionised) for cells in volumes that satisfy the ionisation condition (equation~\ref{eq:new_con}), otherwise they remain fully neutral with zero ionisation fraction.

The left hand side (LHS) of our new ionisation condition
(equation~\ref{eq:new_con}) represents the actual number of escaped
ionising photons being emitted in $dt$. The first term of the right
hand side (RHS) of this condition is the actual number of recombinations
occurring during $dt$ in regions with an ionisation fraction of
$\rm x_{HII}$.  This term then tracks the exact number of recombinations
even from partially ionised regions with $\rm 0 < x_{HII} < 1$. This
recombination term has no effect at early times of EoR when the
universe is completely neutral, but becomes the dominant sink for
ionising photons at late stages of the EoR~\citep{sob14}.

The second term of the RHS of equation~\eqref{eq:new_con} denotes
the total number of neutral hydrogen in the simulation box. At early
EoR stages, the escaped ionising photons (LHS) fights only with the
neutral hydrogen term $\rm (1-x_{HII}) N_H$. As the EoR proceeds,
the neutral hydrogen term becomes less significant, and the
recombinations start to play the leading role in forestalling
reionisation.  Hence this provides a more physically motivated
ionisation condition, in a similar form to the standard ionisation
balance equation.

The condition is clearly time-dependent, unlike the Instantaneous
ionisation model for which the ionisation condition could be evaluated
independently at each time-step.  Thus our new scheme can be sensitive
to the choice of the time-step $dt$ used to perform the integration.
For instance, a larger $dt(dz)$ will result in more ionising photons
and also more recombinations. This then leads to a wrong evolving 
ionisation balance. We have conducted convergence tests to determine
that $\rm dz=0.125$ provides a numerically-converged answer (see Appendix~\ref{sec:contests}).  Our new method thus requires higher computational
cost to evolve the ionisation state forward in time.  We will explore
possible variations of the ionisation condition from equation
~\eqref{eq:new_con} in \S\ref{sec:effects}, to study their impact
on the 21cm power spectrum. 

It turns out, as we will show, that the instantaneous ionisation
condition results in more extended reionisation history, while our
new time-integrated condition yields more sudden reionisation. 
Next, we will investigate their differences in terms of the EoR
history, topology, and the 21cm power spectrum.

\section{Impact on EoR observables}\label{sec:impact}

We use the Full model from ~\citet{hassan16} as our fiducial ``Instantaneous"
reionisation model. This model uses equation\eqref{eq:inst} to
identify the ionised regions with $\fesc = 4\%$ in a large volume
box of $L = 300$~Mpc and a number of cells $N=560^{3}$. This model yields
a maximum halo mass of  $3.56\times 10^{12} M_{\odot}$ and a minimum halo mass 
of $1.28\times 10^{8} M_{\odot}$ at z=6.
We have shown that this model matches various observations of the EoR
including the \citet{planck15} optical depth $\tau=0.066$. Using
the same density field boxes and halo catalogues, we run our new
Time-integrated reionisation model with parameters calibrated against
various EoR key observations (see section~\ref{sec:calibration}),
including the new \citet{planck16} optical depth.  The two models
are tuned to different $\tau_e$ values, but we do our 21cm comparison
at a given neutral fraction since it has been shown that the 21cm
power spectrum shape is more sensitive to the neutral fraction (e.g.
see \citealt{zah07,mesinger07}).  We also verify this later by comparing the Instantaneous model power spectra  at different redshifts for a fixed neutral fraction in figure~\ref{fig:pk21} in \S\ref{sec:21power}.   We thus begin by comparing the
Instantaneous and Time-integrated models' differences in terms of their global
neutral fraction history.

\subsection{EoR ionisation history}

\begin{figure}
\centering
\setlength{\epsfxsize}{0.5\textwidth}
\centerline{\includegraphics[scale=0.35]{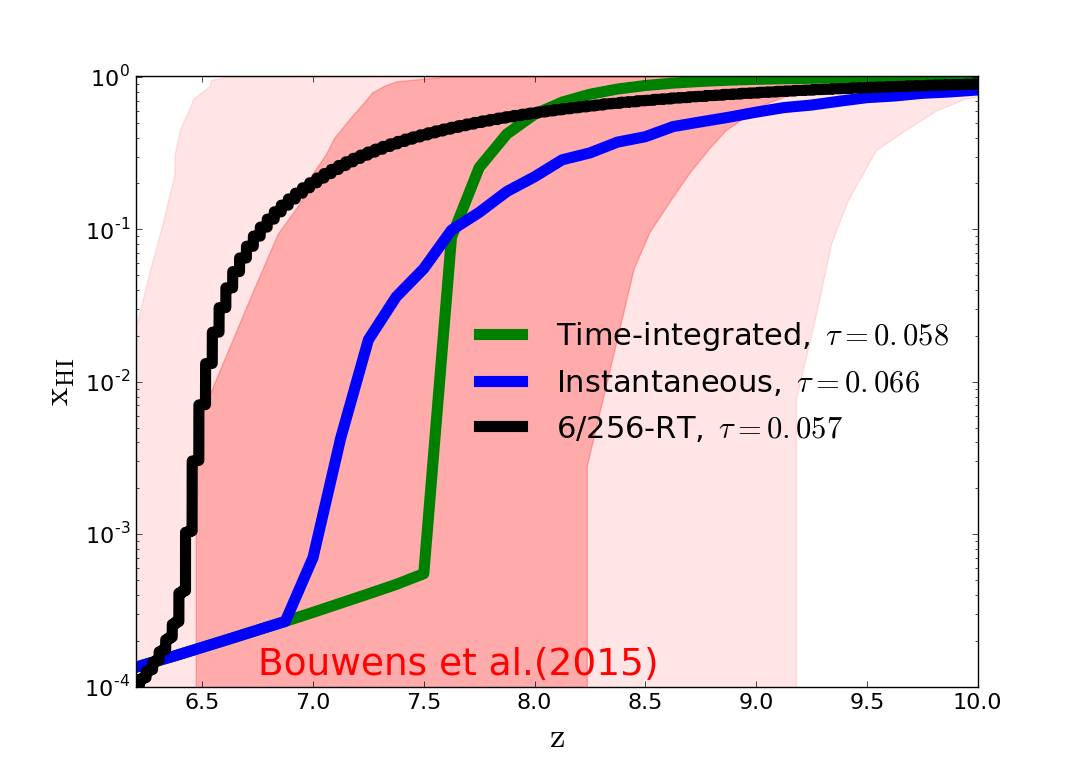}}
\caption{The volume-weighted average neutral fraction evolution as a function of redshift. The Time-integrated EoR model ($\tau=0.058$), the Instantaneous EoR model developed in~\citet{hassan16} ($\tau=0.066$), and the 6/256-RT simulation ($\tau=0.057$) are represented by the green, blue, black lines, respectively. The shaded areas are several quasars and ly-$\alpha$ constraints as compiled by \citet{bouw15a}. Its quite clear that all models are consistent with the observational constraints by \citet{bouw15a}. Differences between models are explained in the text. }
\label{fig:eorhistory}
\end{figure} 

Figure~\ref{fig:eorhistory} shows the global reionisation history
produced by our two fiducial models, Instantaneous and Time-integrated, compared
to the neutral fraction constraints obtained by \citet{bouw15a} via
a compilation from various observables.   We immediately see that
the green line showing the new time-integrated ionisation condition
shows a more sudden transition from fully ionised to fully neutral.  Meanwhile, the blue line
from our old instantaneous condition results in a more extended
reionisation epoch.  Nonetheless, in both cases, reionisation occurs
in our two models within observational constraints (light-red shaded
areas).  It is perhaps worth noting that, unlike a few years ago
when the canonical redshift for the end of reionisation was regarded
as $z\sim 6$, current constraints from both observations and models
favors the end of reionisation to occur at $z\sim 7$ or perhaps a bit
higher.

This plot already shows that accounting for the neutral gas through
comparing the number of neutral atoms and ionising photons
(equation~\ref{eq:new_con}) versus comparing instantaneous rates
(equation~\ref{eq:inst}) has a significant impact on the reionisation
history. The Time-integrated model is qualitatively more compatible with the
picture that has been suggested by recent \citet{planck16} constraints
that favours sudden EoR scenarios. We emphasize the fact that if
we tune the Time-integrated model optical depth to match the Instantaneous model
optical depth ($\tau = 0.066$,~\citealt{planck15}), the Time-integrated model
will require higher $\fesc$ and shift reionisation towards higher
redshifts, but nevertheless the reionisation history shape will
remain sudden as shown later in figure~\ref{fig:xHItaucons}. We
will come back to this point later in \S\ref{sec:tau_constraints}. 
However, when using the same parameters, the reionisation in the time-integrated model is delayed 
by $\Delta z \sim 0.8$ as compared with that in the instantaneous model.

From figure~\ref{fig:eorhistory}, we also see that both models
Instantaneous and Time-integrated reionise the universe earlier than the 6/256-RT
simulation. As discussed before in ~\citet{hassan16}, the small box
size (6 h$^{-1}$ Mpc) of 6/256-RT does not capture the large scale
fluctuations that give rise to the most massive halos that provide
a significant fraction of reionising photons.  Hence at a fixed
optical depth, it is expected that the 6/256-RT might reionise the
universe much later than the Time-integrated model due to the box size
limitations.

\begin{figure*}
\setlength{\epsfxsize}{0.5\textwidth}
\centerline{\includegraphics[scale=0.55]{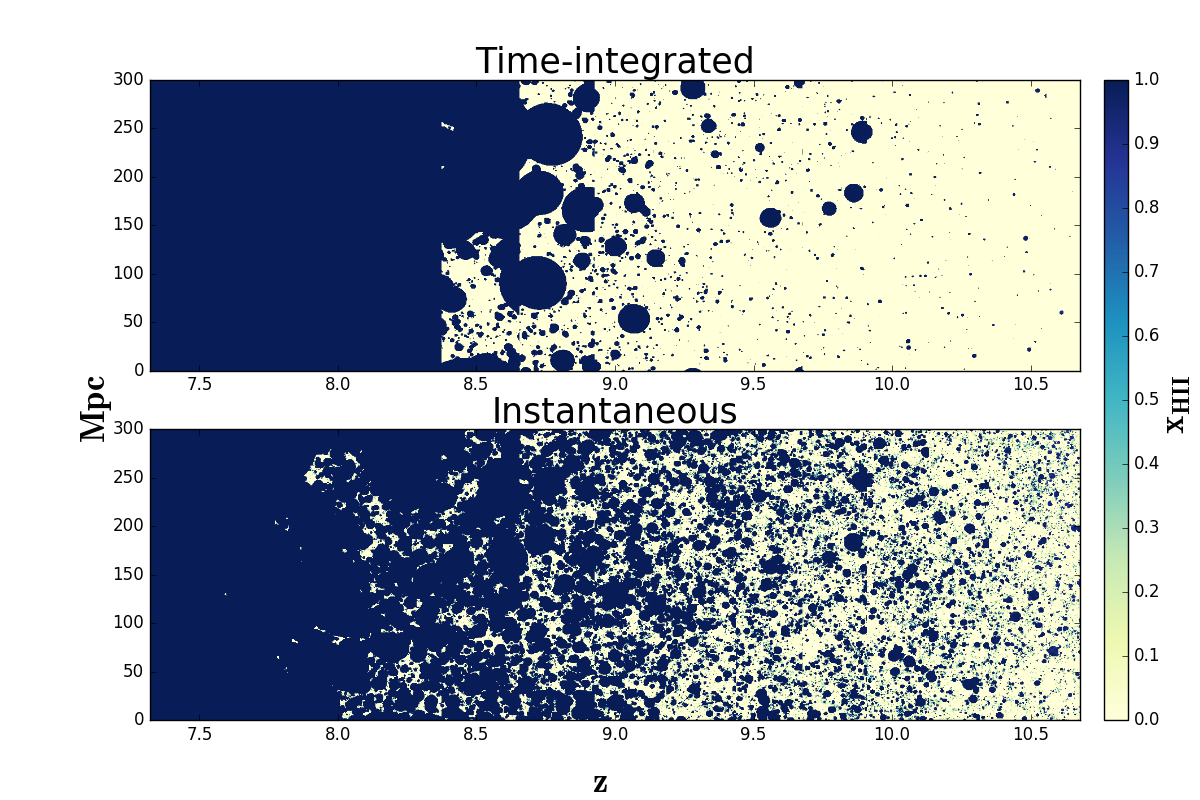}}
\caption{Evolving maps of the neutral fraction from the Instantaneous and Time-integrated models. Time-integrated EoR model produces large HII bubbles and reionises the universe very rapidly, indicating a sudden EoR scenario. Instantaneous EoR model yields small HII bubbles and reionises the universe very late, leading to an extended EoR scenario.}
\label{fig:lightcones}
\end{figure*}

An informative way to examine these models is by viewing light
cones, as shown in Figure~\ref{fig:lightcones}. These have been
constructed by projecting the ionisation state within the simulation
volume along a specific line-of-sight, evolving with redshift.
Figure~\ref{fig:lightcones} confirms that our previous model (the
Full model of ~\citealt{hassan16}) produces a more extended EoR
scenario that corresponds to an early onset and a very late end
with a duration of $\Delta z \sim 10$. Unlike the Instantaneous model,
figure~\ref{fig:lightcones} also shows that our new model
(equation~\ref{eq:new_con}) yields a sudden reionisation scenario
where the EoR starts very late (once $\rm x_{HI} < 1.0$ ) and ends (when $\rm x_{HI}$ drops below 10$^{-3}$) very quickly within a
duration of $\Delta z \sim 4$.  More strikingly, it also shows that the Time-integrated EoR model produces
larger ionised bubbles while the Instantaneous model yields many ionised
bubbles of smaller sizes. This can further be quantified by studying
their differences in terms of the ionisation field power spectra.

\begin{figure*}
\setlength{\epsfxsize}{0.5\textwidth}
\centerline{\includegraphics[scale=0.6]{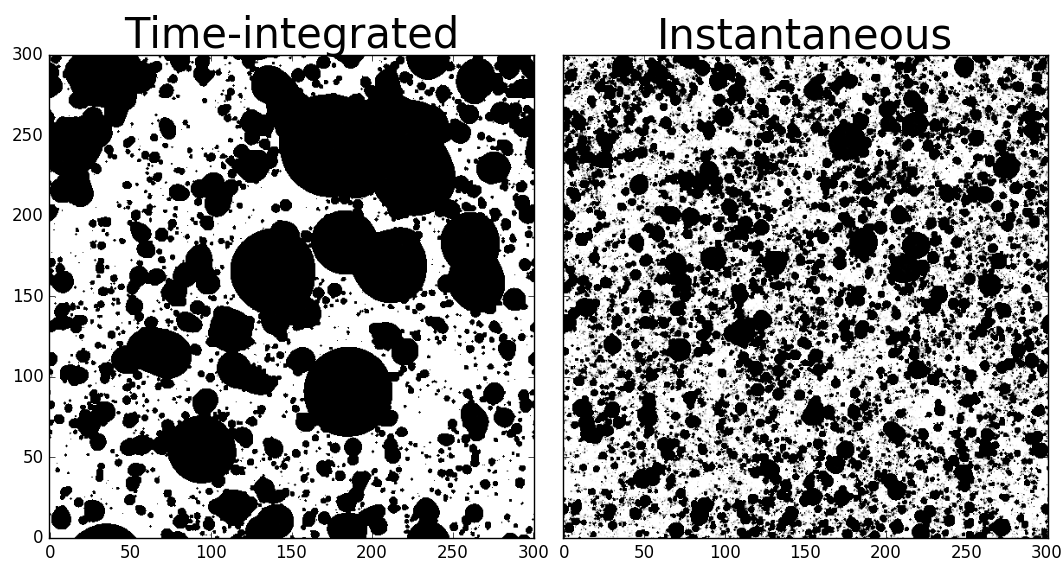}}
\caption{Slice of the ionistion box of a size $\rm 300\, \times\, 300\, \times\, 0.535\, Mpc^{3}$ from the Instantaneous and Time-integrated models and at $\rm x_{HI}\sim 0.5$. White and black represent neutral and ionised regions respectively. The self-shielded regions in ionised mediums are absent due to the binary structure of the excursion set formalism (assigning 1 to ionised and 0 to neutral cells in regions that satisfy the ionised condition) along with the large cell size of $\sim$ 0.5 Mpc.}
\label{fig:ion_maps}
\end{figure*}

\subsection{EoR topology}   

It is useful to compare the models at a specific neutral fraction,
since this best illustrates the difference in topology.
Figure~\ref{fig:ion_maps} compares our Instantaneous and Time-integrated models
in terms of their ionisation maps when the EoR is half-way through,
i.e.  with a globally-averaged $\rm x_{HI} \sim 0.5$. These ionisation
maps show the spatial distribution of the large and small ionised
bubbles (black regions) over 300 Mpc scales. 
However, the excursion set-formalism 
with its binary structure (fully ionised or fully neutral), along with the large cell size of $\sim$ 0.5 Mpc, prevents these models' maps to display
the self-shielded regions in the ionised mediums as seen in figures \ref{fig:ion_maps} and \ref{fig:lightcones}.
The presence of these self-shielded regions does not affect the 21cm power spectrum but rather lower the ionisation fraction at intermediate densities ($\Delta = 5 - 10$) as previously shown (see figure (10) in \citealt{hassan16}). We also have shown in \citet{hassan16} that including the sub-clumping effect on scales below our cell size ($\sim$ 0.5 Mpc) have a minimal effect on the expected signal (see comparison between {\bf Full} and {\bf NoSubClump} models in \citealt{hassan16} for more details).

From Figure~\ref{fig:ion_maps}, we see  that the Instantaneous model produces
many small \ion{H}{ii} bubbles more uniformly distributed across
the ionisation map. This shows that the EoR in the Instantaneous model
proceeds from small scales, and the ionising photons are able to
reionise locally everywhere.  This is because the instantaneous
ionisation rate can easily exceed the recombination rate (see
equation~\eqref{eq:inst}) on small scales when neglecting the local
neutral hydrogen content.

In contrast, the Time-integrated model ionisation map shows very large
\ion{H}{ii} bubbles.  This may be explained by interpreting the
Time-integrated model ionisation condition (equation~\ref{eq:new_con}).
As noted earlier, at high redshifts when the universe is neutral
($\rm x_{HII} \sim 0$), the recombination term can be neglected.
In this case, the Time-integrated model only compares the escaped ionising
photons with the total number of neutral hydrogen atoms.  This
condition dominates until the region becomes partially ionised. At
that point, recombinations will start to occur because of the nonzero
ionisation fraction $\rm x_{HII}$, but still this region is now
less neutral, which will allow more rapid ionisation. However,
different forms of ionisation condition yield very different HI fluctuations (see \S\ref{sec:effects}).   In general,
the sources and sinks are occurring within the regions that are
densest and thus contain the most number of neutral hydrogen atoms.
The high density causes the ionising photons to be ineffective at
ionising local regions, until such time as significant ionisations
happen, which then rapidly ionise the surrounding regions.

\begin{figure*}
\setlength{\epsfxsize}{0.5\textwidth}
\centerline{\includegraphics[scale=0.45]{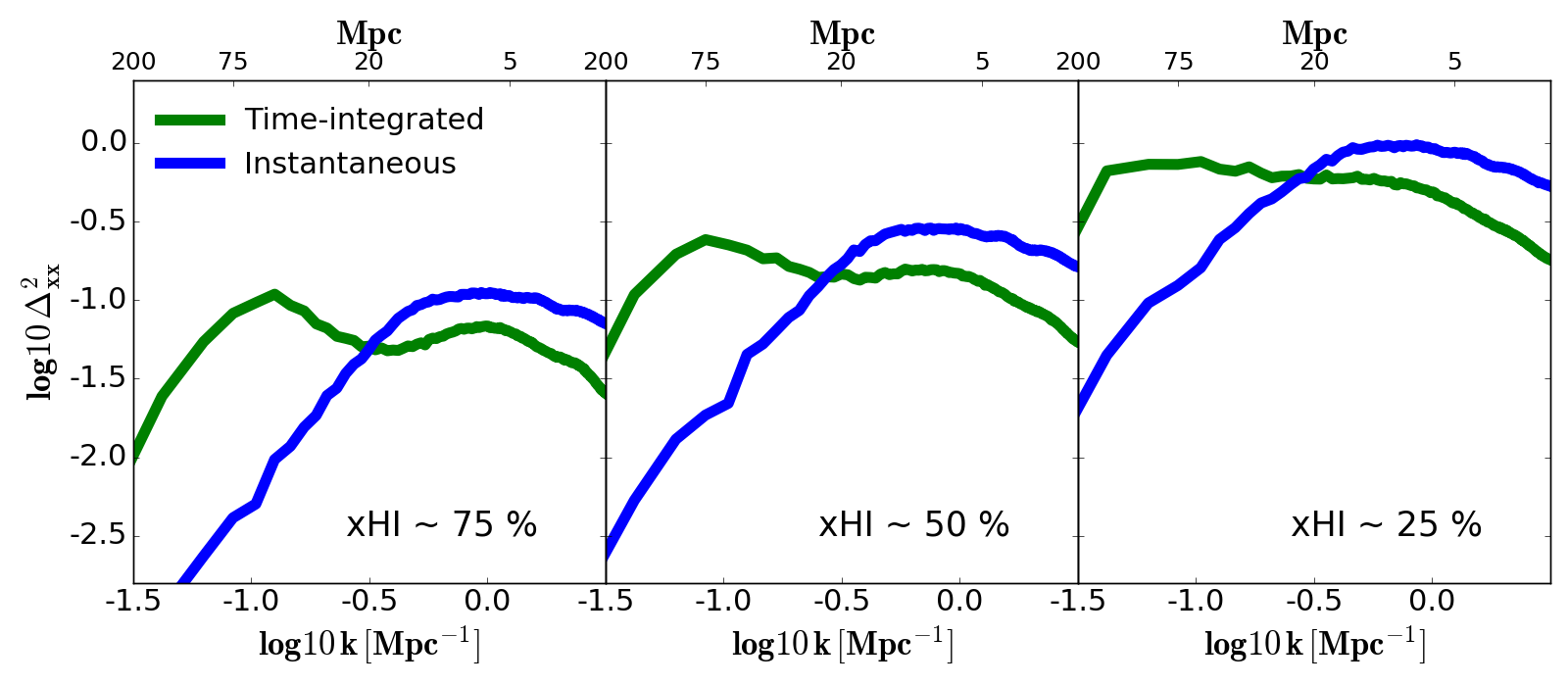}}
\caption{Ionization field power spectrum comparison between the Instantaneous (blue) and Time-integrated (green) models at different stages of reionisation ($\rm x_{HI}\, \sim\, 25\%,\, 50\%,\, 75\%$.)}
\label{fig:pkxi}
\end{figure*}

Figure~\ref{fig:pkxi} shows the ionisation field power spectra of
our fiducial models at different stages of reionisation when the
universe is 25\%, 50\% and 75\% reionised. These neutral fractions correspond to z=8.0,8.75,9.5 
and z= 7.75,8.0,8.25 as obtained by the Instantaneous and Time-integrated models respectively. This is computed as
follows: $\mathrm{\Delta^{2}_{xx} \equiv k^{3}/|(2\pi^{2}\, V) < |
x_{HII}|^{2}>/x^{2}_{HI}}$~\citep{hassan16}.

The Time-integrated model produces more power on large scales by 1-1.2 order
of magnitude and less power on small scales by a factor of 2-3, at
fixed ionisation fraction, as compared to the Instantaneous model.  This
is consistent with the qualitative impression from the ionisation
maps in Figure~\ref{fig:ion_maps}.  We further see that the large
scale ionisation power spectrum, obtained by the Time-integrated EoR model,
peaks at $\sim$ 75Mpc which corresponds to the characteristic size
of the \ion{H}{ii} bubbles as seen in the \ion{H}{ii} maps in
figure~\ref{fig:ion_maps}.

The difference particularly on large scales is substantial, which
shows the importance of accounting for the existing neutral hydrogen
content in the ionisation condition (i.e. the second term of
equation~\eqref{eq:new_con}).  Since the fluctuations in the
ionisation field drive the 21cm brightness temperature, we expect
to see similar differences in the 21cm power spectra, which we
examine next.

\subsection{The 21cm power spectrum}\label{sec:21power}

\begin{figure*}
\setlength{\epsfxsize}{0.5\textwidth}
\centerline{\includegraphics[scale=0.45]{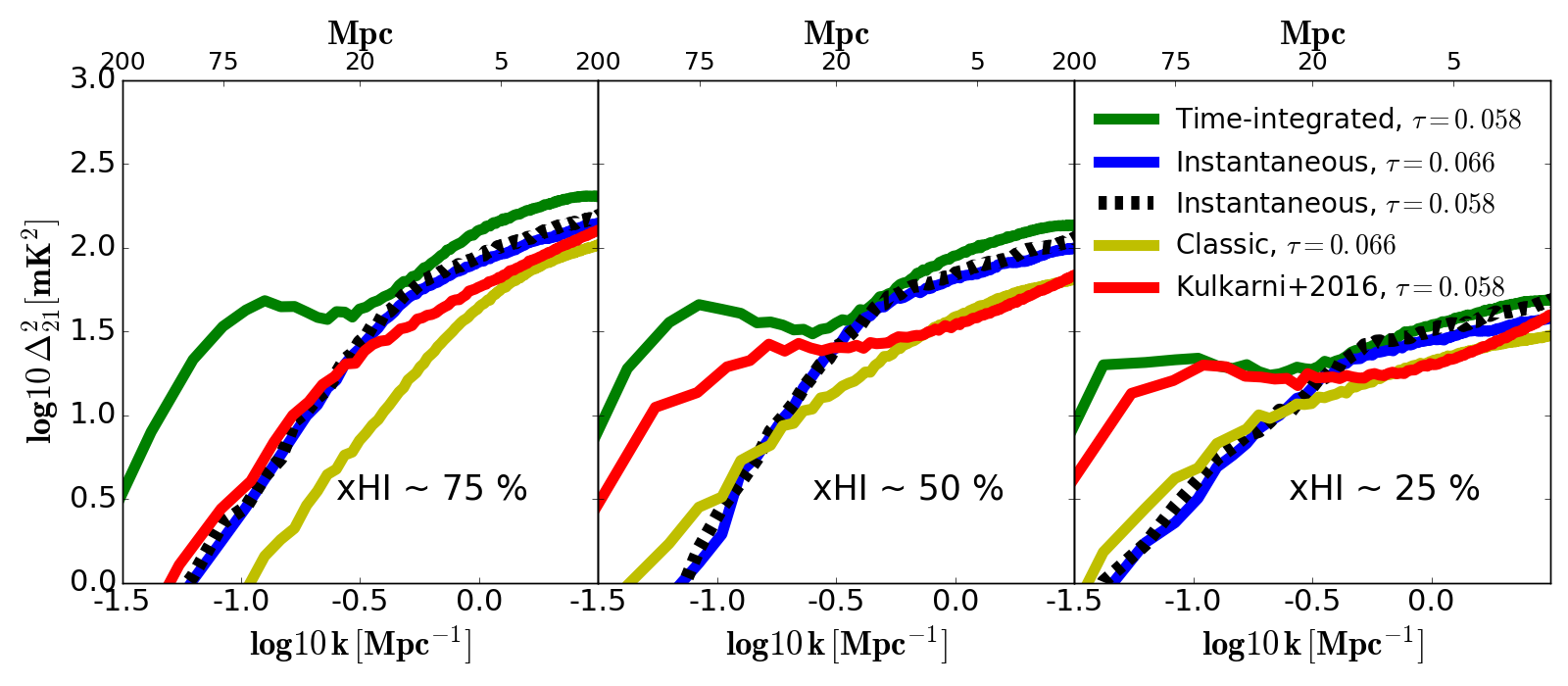}}
\caption{The 21cm power spectrum comparison between the Instantaneous (blue) and Time-integrated (green) models at different stages of reionisation ($\rm x_{HI}\, \sim\, 25\%,\, 50\%,\, 75\%$). We compare our 21cm power spectra with the Very Late model (red) by similar semi-numerical method \citet{kul16} that is calibrated to match Ly$\alpha$ and CMB data. Although our Time-integrated and \citet{kul16} models correspond to different redshifts, but nevertheless the shape of the 21cm spectrum is similar, particularly at the intermediate and late stages of reionisations. We also show our Classic EoR model (yellow) from~\citet{hassan16} which adopts the standard efficiency parameter approach similar to \citet{kul16} model. Our Classic and \citet{kul16} models produce similar power on small scales. The Instantaneous model, whether it is tuned to \citet{planck16} $\tau$ (black dashed) or \citet{planck15} $\tau$ (solid blue), always produces the exact 21cm power spectrum at fixed neutral fractions, regardless of the density field contribution from different redshifts. Models are tuned to the two recent {\it Planck} optical depth values as quoted in the legend.}
\label{fig:pk21}
\end{figure*}
Using the ionisation fields of these models, we now compute our EoR key observable which is the 21cm power spectrum. Assuming that the spin temperature is much higher than the CMB temperature, the 21cm brightness temperature takes the following form:
  \begin{equation}\label{21cm}
\delta T_{b} (\nu) = 23\mathrm{ x_{HI} }\Delta \left( \frac{\Omega_{b} h^{2}}{0.02} \right) \sqrt{  \frac{1+z}{10} \frac{0.15}{\Omega_{m}h^{2}}} \left( \frac{H}{H + d v/dr} \right) \mathrm{mK},
\end{equation}
where $d v/dr$ is the comoving gradient of the line of sight component of the comoving velocity. Using this equation, it is straightforward to create the 21cm brightness temperature boxes from which we compute the 21cm power spectrum as follows: $\mathrm{\Delta^{2}_{21} \equiv k^{3}/(2\pi^{2}\, V) < | \delta T_{b} (k)  |^{2}_{k} >}$.

 We first verify that the 21cm power spectrum is primarily
sensitive to the global ionisation fraction, while the density field
evolution is secondary.  Note that the 21cm fluctuations traces
those of the density field only at early times when the universe
is almost neutral. Here we quote results for the Instantaneous model,
but we expect that this is also valid for other models such as
our Time-integrated model. We tune the Instantaneous model to
\citet{planck16} optical depth ($\tau=0.058$) which yields $\rm
x_{HI} = 0.75,0.54,0.277$ at $\rm z=8.75,8.00,7.25$ respectively.
We then re-tune the model to the \citet{planck15} optical depth
($\tau=0.066$) (similar to our previous {\bf Full} model in
\citet{hassan16}) to obtain $\rm x_{HI} = 0.73,0.53,0.25$ at $\rm
z=9.5,8.75,8.0$ respectively. We now compare their difference in
the 21 power spectrum at these different redshifts for a fixed
neutral fraction in figure~\ref{fig:pk21}. Comparing the solid blue
line with black dashed line, we find the the Instantaneous model
produces the exact 21cm power spectrum at a fixed neutral fraction,
irrespective of the density field evolution at different redshifts.
Hence we will compare different models at similar neutral fractions,
not similar redshifts.

Figure~\ref{fig:pk21} shows the 21cm power spectrum of the Instantaneous
and Time-integrated models at neutral fractions of  25\%, 50\% and 75\%.
Mimicking the ionisation field power spectrum, the Time-integrated model
produces more power on large scales by 1-1.2 order of magnitude at
fixed ionisation fraction, as to that of the Instantaneous model.
Likewise, the Time-integrated model also produces slightly more power on
small scales by a factor of 1.2-1.5 as compared with the Instantaneous
EoR model. This difference is less than when comparing the ionisation
field power spectra, which comes from the contribution of the density
field to the 21cm power spectrum -- small regions with high local
density (high recombinations) remain neutral, and hence they do not
contribute much to the small-scale fluctuations in 21cm power.

We also compare our 21cm power spectra to a similar semi-numerical
model by~\citet{kul16} that has been calibrated with Ly$\alpha$ and
CMB data. The semi-numerical models by \citet{kul16} adopts the standard 
efficiency parameter ($\zeta$) approach similar to our Classic EoR model (yellow in fig~\ref{fig:pk21})
from ~\citet{hassan16}. We choose to compare with the Very Late model in~\citet{kul16} (red in fig~\ref{fig:pk21})
that is tuned to match the \citet{planck16} optical depth, consistent
with the optical depth produced by our Time-integrated model. The ionisation
histories of \citet{kul16} and our Time-integrated model are very different
even though they obtain $\sim 50\%$ neutral fraction at the same redshift.
For instance, our Time-integrated model produces $\rm x_{HI} =
0.77, 0.57, 0.25$ at z=8.25,8.0,7.75 whereas \citet{kul16} model
finds  $\rm x_{HI} = 0.84, 0.59, 0.42$ at z=10.0,8.0,7.0. Regardless
of this difference in these models' reionisation histories, their
21cm power spectra are generally similar. We find that both models
produce a similar shape of the 21cm power spectrum particularly
during the intermediate and final stages of reionisation. The minor
difference in their amplitudes is due to using our $\rion$-$\rrec$ versus the standard $\zeta$ approach.
This can be clearly seen when comparing our Classic EoR model with  \citet{kul16} model in figure~\ref{fig:pk21}.
We see both models produce the same power on small scales while their difference on large scales might be from 
the difference in the density field and neutral fractions. This confirms our previous findings that using the non-linear
ionisation power, via our $\rion$-$\rrec$ approach, boosts the 21cm power spectrum as compared to models adopting the standard efficiency parameter method (Classic and \citet{kul16} models).  
 
This shows that our Time-integrated model, that is calibrated to match various
EoR key observables, produces similar 21cm power spectrum as obtained
by other semi-numerical models that have been calibrated to match
Ly$\alpha$ and CMB data. The future 21cm observations might be able to discriminate between
these models' power spectra. 
 
In summary, we have compared the Instantaneous and Time-integrated models in terms
of their EoR history, topology, and their 21cm power spectra.  We 
have found that the Time-integrated model produces large HI bubbles while the Instantaneous model produces more small HI bubbles. The Time-integrated model yields a large scale 21cm/Ionization power spectrum that is higher by 1 order of magnitude as compared with the Instantaneous model.
We have seen that the ionisation condition (equation~\ref{eq:new_con}) results in large \ion{H}{ii} bubbles
which boost the amount of power on large scales. The comparison presented here  aims to summarize the differences found between our new (Time-integrated) and previous (Instantaneous) models. However, previous works by ~\citet{zah11} and ~\citet{maj14} have shown that semi-numerical simulations agree with radiative transfer simulations in terms of their ionization fields and 21cm power spectra. We leave for future work whether this new model matches radiative transfer simulations.

\begin{figure}
\setlength{\epsfxsize}{0.45\textwidth}
\centerline{\includegraphics[scale=0.42]{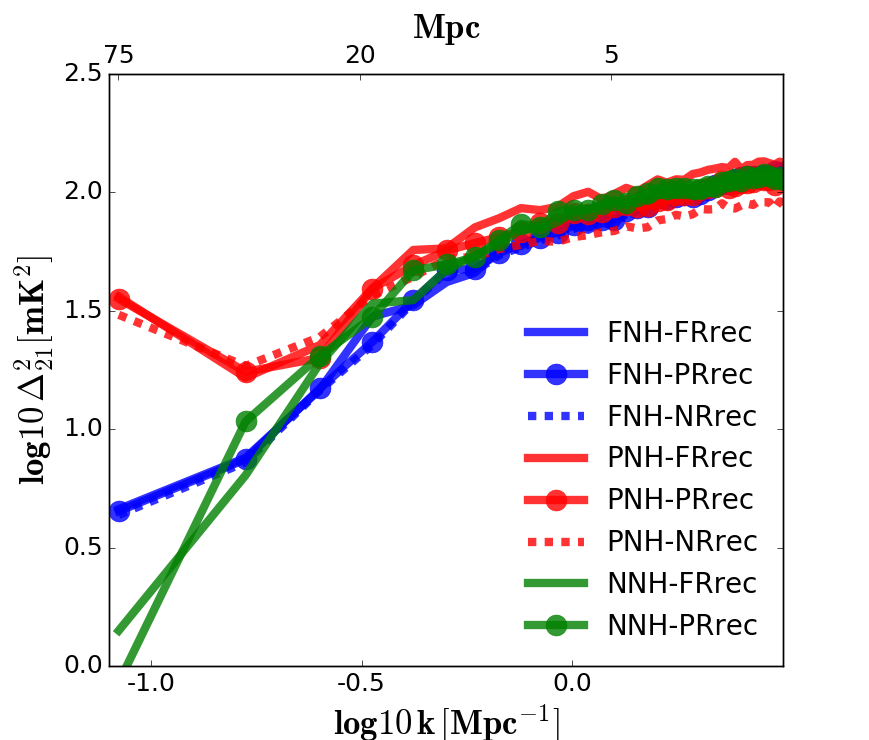}}
\caption{The 21cm power spectrum comparison for different physical assumptions at $\rm x_{HI}\, \sim\, 50\%$. Different colors represent different ways to treat the local fluctuations in the neutral hydrogen density while different line styles corresponds to different recombinations terms, as explained by the legend and text. It is evident that the recombinations are subdominant in determining the large scale 21cm power spectrum. It is also shown models (red), that track the neutral fraction from partially ionised regions, yield a very high 21cm power spectrum on large scales.   }
\label{fig:pk21models}
\end{figure}

\section{Model assumption effects on the 21cm power spectrum}\label{sec:effects}

The large differences in the 21cm power spectrum (figure~\ref{fig:pk21})
between the Instantaneous and Time-integrated models show that the 21cm power
spectrum is highly sensitive to the physical assumptions used.
There are two main differences between these models:  First, the
ionisation condition now accounts for the number of hydrogen atoms,
and second, the recombination is now done accounting for partial
ionisation.  We believe our new model is more physically-motivated
and realistic, but we would like to understand exactly how these
changes individually impact the 21cm power spectra.

We therefore consider the various possible combinations between
$\rrec$ and $\rm N_{H}$ to create several models with different
ionisation condition.  We also consider models neglecting recombination
altogether to analyse the impact of recombinations as we did in
\citet{hassan16}, only with our new Time-integrated model.

More specifically, we keep the LHS of equation~\eqref{eq:new_con}
($\rion$ term) same and vary the integrand of RHS integrals, namely $\rm {I\!R}_\mathrm{rec,V}$ and $\rm N_{HI,V}$, to create the following models:
\begin{itemize} 
\item Full-NH-Full-Rrec ({\bf FNH-FRrec}): $\rm  \rrec \,+ N_{H}$.
\item Full-NH-Partial-Rrec ({\bf FNH-PRrec}): $\rm x_{HII} \,\rrec\, + N_{H}$.
\item Full-NH-No-Rrec ({\bf FNH-NRrec}): $\rm N_{H}$.
\item Partial-NH-Full-Rrec ({\bf PNH-FRrec}): $\rm \rrec\, + (1-x_{HII})\,N_{H}$.
\item Partial-NH-Partial-Rrec ({\bf PNH-PRrec}): $\rm x_{HII}\,\rrec\, +  (1-x_{HII})\,N_{H}$.
\item Partial-NH-No-Rrec ({\bf PNH-NRrec}): $\rm (1-x_{HII})\,N_{H}$.
\item No-NH-Full-Rrec ({\bf NNH-FRrec}): $\rm  \rrec\,$.
\item No-NH-Partial-Rrec ({\bf NNH-PRrec}): $\rm x_{HII}\,\rrec\,$.
\end{itemize}

\begin{table*} \LARGE
 \scalebox{0.75}{\begin{tabular}{ || l || c || c||}\hline
   {\bf Model Class }& {\bf Recombination term} & {\bf Neutral Hydrogen term} \\ \hline \hline
    {\bf Full-NH-Full-Rrec     ({\bf FNH-FRrec})} & $\rm \rrec$  & $\rm  N_{H} $\\ \hline 
    {\bf Full-NH-Partial-Rrec  ({\bf FNH-PRrec}) } & $\rm x_{HII}\, \rrec$  & $\rm  N_{H} $\\ \hline 
    {\bf  Full-NH-No-Rrec ({\bf FNH-NRrec})} & 0 & $\rm  N_{H} $\\ \hline 
    {\bf Partial-NH-Full-Rrec ({\bf PNH-FRrec})} & $\rm \rrec$  & $\rm (1-x_{HII})\, N_{H} $\\ \hline 
    {\bf  Partial-NH-Partial-Rrec ({\bf PNH-PRrec}) } & $\rm x_{HII}\, \rrec$  & $\rm (1-x_{HII})\, N_{H} $\\ \hline 
    {\bf Partial-NH-No-Rrec ({\bf PNH-NRrec}) } & 0 & $\rm (1-x_{HII})\, N_{H} $\\ \hline 
    {\bf No-NH-Full-Rrec ({\bf NNH-FRrec}) } & $\rm \rrec$  & 0\\ \hline 
    {\bf No-NH-Partial-Rrec ({\bf NNH-PRrec})} & $\rm x_{HII}\, \rrec$  & 0\\ \hline 
\end{tabular}}
\caption{Summary of models considered in section~\ref{sec:effects} for the 21cm power spectrum comparison in figure~\ref{fig:pk21models} from different physical assumnptions and treatment to the integrands of RHS integrals in the time-integrated ionisation condition, equation~\eqref{eq:new_con}. }\label{models_tab}
\end{table*}

The Time-integrated model is represented here by ionisation condition of
{\bf PNH-PRrec} whereas the Instantaneous model uses that of {\bf
NNH-FRrec}. The others are variants on these  as summarized in Table~\ref{models_tab}.  To illustrate the
differences in the 21cm power spectrum, we use the same density
field and halo catalogues generated within a simulation run of a
box size 75 Mpc and N = 140$^{3}$. We have shown previously (Figure
8 in~\citealt{hassan16}) that the numerical volume convergence of
our simulated 21cm power spectrum is excellent at all redshifts
down to a box size of 75 Mpc, hence, we expect the same 21cm power
spectrum for larger simulation volumes. The reionisation history
produced by these models vary, so as before we choose to make our
21cm power spectrum comparison at a fixed neutral fraction.

Figure~\ref{fig:pk21models} shows the 21cm power spectrum produced
by different models at 50\% neutral fraction as explained above.
First, we see that all these variants result in virtually the same
21cm power spectrum on small scales. This reiterates our previous
finding in ~\citet{hassan16} that using a non-linear ionisation
rate $\rion$ boosts the 21cm power spectrum by a similar amount
regardless of whether one accounts for recombinations or not, and
further shows that accounting for the neutral hydrogen atoms does
not alter this conclusion.

More significant differences are evident at large scales for the
21cm power spectrum. Models starting with full $N_H$ ({\bf FNH-})
produce 21cm power spectra with the same shape and amplitude on
all scales (i.e. all the blue lines overlap), and likewise for
models with partial or no $N_H$ ({\bf PNH-} and {\bf NNH-}).  This
demonstrates that recombinations are subdominant for determining
the large-scale 21cm power spectrum.  It is clear that the {\bf
NH} term plays a major role in boosting/suppressing the large scale
21cm power spectrum. This means that semi-numerical models must
carefully account for the local number density of neutral hydrogen
for a proper prediction of the expected signal.

From figure~\ref{fig:pk21models}, we see the clear trend that models
that do not account for the existing neutral hydrogen atoms ({\bf
NNH-} models such as the Instantaneous model) have lower 21cm power
spectrum on large scales. Furthermore, models that use the total
number of hydrogen atoms ({\bf FNH-} models) at each time-step
regardless of the ionisation fraction show 21cm power spectra that
is slightly higher on large scales as compared to {\bf NNH-} models. This is due to
the presence of weak HI fluctuations by following only the density field  ($\rm \bf N_{HI} \sim \Delta$).  However,
models, that use the ionisation history of cells to track the neutral
hydrogen atoms from partially ionised regions ({\bf PNH-} models
such as the Time-integrated model), show a very high 21cm power spectrum
on large scales as opposed to the {\bf NNH-} and {\bf FNH-} models. 
This comes from the fact that the the {\bf PNH-} models account for a strong HI fluctuations by following the density field ($\rm \bf N_{HّّI}  \sim \Delta$) and ionisation field ($ \rm \bf N_{HّّّI} \sim x_{HII}$) both.
This shows that, at given neutral fraction, the large scale 21cm
power spectrum is highly influenced by the way in which we account
for the fluctuations in the local neutral hydrogen density.

In the next section, we will discuss the calibration of the Time-integrated
model against various EoR key observables and test how well the
ongoing/upcoming 21cm observations will further constraints our
free parameters.

\section{Model calibration}\label{sec:calibration}

We now focus on our favoured Time-integrated reionisation model, which
includes all our new physics implementations.  Previously, the
parametrization of $\rion$ (equation~\ref{eq:nion}) was obtained
from our small-volume high-resolution radiative transfer hydrodynamic
simulation (6/256-RT).  However, the small volume of this simulation
makes it subject to uncertainties since, as we saw in Figure~\ref{fig:eorhistory},
the ionisation history of this simulation is significantly delayed
by its small volume.  Here, we adopt a more general form for $\rion$,
and determine whether existing EoR measurements can calibrate our
source model, and thereby provide constraints on the nature of
reionising sources.

To this end, we here consider a more generalized model with the
following three free parameters:
\begin{itemize}
\item $\fesc$ is the volume-averaged photon escape fraction. 
\item $\aion$ is the ionising emissivity amplitude, which scales the amount of ionising emissivity ($\rion$) equally across the halo mass range at a given redshift.
\item $\cion$ is ionising emissivity-halo mass power dependence, which quantifies the $\rion$-$M_{h}$ slope.
\end{itemize}

We will constrain these three parameters against various EoR
observations and compare with the values found from fitting to the
6/256-RT simulation, using a Bayesian MCMC approach.
We choose these parameters to explore since they are
most closely related to the emission characteristics of the source
population.  We ignore $B_{\rm ion}$ which is related to how
photoionisation suppresses low-mass galaxy growth, and $D_{\rm ion}$
because it is not physically obvious why ionisation rate of a given
halo should have a strong redshift dependence.  While it would be
better to simply let all these parameters vary, even doing an MCMC
over this 3-D space is already computationally challenging since
it requires doing full runs for each sampling, and increasing the
dimensionality quickly makes the computational requirements
intractable.

Recall that the Time-integrated model identifies the ionised regions
using a time dependent ionisation condition (equation~\ref{eq:new_con}),
that tracks the exact recombinations and neutral hydrogen atoms by
following the reionisation history. The reionisation history, in
this model, is numerically well converged for $\Delta z \leq 0.125$.
With these requirements, the model becomes more computationally
expensive to run, but nevertheless, is feasible for independent
large volume runs.  However, sampling the full MCMC space requires
at least $\sim\, 10^{6}$ simulation realizations, which becomes
infeasible.  Hence we precompute a grid of models spanning the full prior
space, and then do a trilinear interpolation to
obtain the observables for any given parameter combination. 
This sacrifices some accuracy but makes the computation feasible.

We note that \citet{Greig15} developed an analysis pipeline, {\sc
21cmmc}, that directly links their semi-numerical model 21CMFAST
~\citep{mesinger11} to a Bayesian routine {\sc cosmoHammar}~\citep{aker13}
to constraining their free parameters. However, their ionisation
condition did not include recombinations through the time integral
method which they developed in \citet{sob14}, and instead used a
standard efficiency parameter ($\zeta$) approach.  Along with lower
resolution of $\sim\,$ 2 Mpc, these simplifications enabled them
to run their semi-numerical model fully within an MCMC scheme.

\subsection{Parameter estimations pipeline}\label{sec:pipeline}

We choose a cell size of 0.375 h$^{-1}$ Mpc and a box size $L= 75$
Mpc, giving $N=140$ cells per side.  We precompute a grid of $25
\times 25 \times 25$ runs outputting the predicted observables
for our models, uniformly sampling our selected prior range for our
parameters of ($\fesc,\,\rm log10(A_{ion}),\, C_{ion}$) =
[(0,1),(37,44),(-1,2)].  This gives a total of 15,625 simulation
independent realizations, which we interpolate inside the MCMC
search process.

We have tested our parameter constraints using two different Bayesian inference
tools {\sc Multinest}~\citep{feroz09} and {\sc emcee}~\citep{foreman13}.
We have found the same parameter estimates using these two different
codes, and hence, our presented parameters estimation here
appear to be robust to variations in the algorithm used.

\begin{table*} \LARGE
 \scalebox{0.75}{\begin{tabular}{ || l || c || c || c ||}\hline
   {\bf EoR Constraint }& ${\rm\bf \fesc}$ & ${\rm \bf log10(A_{ion})}$ & ${ \rm\bf C_{ion} }$\\ \hline \hline
    {\bf \citet{bouw15} SFR all at z=8,7,6} & 0.51$^{+0.33}_{- 0.34}$  & 39.61$^{+0.18}_{- 0.16}$  & 0.45$^{+0.08}_{- 0.09}$ \\ \hline 
   {\bf \citet{planck16} optical depth $\tau$} & 0.46$^{+0.36}_{- 0.32}$ & 39.08$^{+1.39}_{-1.44}$  & 0.28$^{+0.76}_{- 0.88}$ \\ \hline
	 {\bf \citet{bec13} ionising emissivity $\bf \rm N_{ion}$ at z=4.75} & 0.51$^{+0.34}_{- 0.34}$  & 39.68$^{+0.93}_{- 1.36}$  & -0.12$^{+0.78}_{- 0.62}$ \\ \hline    
    {\bf ALL = SFR+$\tau$+$\rm N_{ion}$} & 0.25$^{+0.26}_{- 0.13}$  & 39.62$^{+0.17}_{- 0.18}$  & 0.44$^{+0.09}_{- 0.09}$ \\ \hline       \hline  
    {\bf Values obtained from fitting to 6/256-RT} &  -   & 40.03  & 0.41 \\ \hline
\end{tabular}}
\caption{Summary of our parameter estimations from individual and combined set of observations, as well as from matching to the 6/256-RT simulations.}\label{constraints_tab}
\end{table*}

We here present the results obtained by using the {\sc emcee} python
package. We use 100 random walkers initialised around the maximum
likelihood. For each walker, we sample 10,000 chains from the
likelihood after 500 initial burn-in chains to achieve convergence.
This makes a total of 1,000,000 samples which is sufficient to explore
the whole parameter space.

\subsection{EoR key observables constraints}

We constrain our three free parameters to the following observations:
\begin{enumerate}
\item The dust-corrected star formation rate density integrated down to $\rm M_{AB} = -17$ by \citet{bouw15} at the following redshifts:
\begin{itemize}
\item z $\sim$ 6: log10(SFR) [M$_{\odot}$ Mpc$^{-3}$ yr$^{-1}$] = -1.55 $\pm$ 0.06.
\item z $\sim$ 7: log10(SFR) [M$_{\odot}$ Mpc$^{-3}$ yr$^{-1}$] = -1.69 $\pm$ 0.06.
\item z $\sim$ 8: log10(SFR) [M$_{\odot}$ Mpc$^{-3}$ yr$^{-1}$] = -2.08 $\pm$ 0.07.
\end{itemize}
\item The \citet{planck16} integrated optical depth to Thomson scattering: $\tau = 0.058 \pm 0.012$.
\item The \citet{bec13} ionising emissivity density measurements from Ly$\alpha$ data at $z=4.75$: $\rm \dot{N}ion$ [10$^{51}$ photons s$^{-1}$ Mpc$^{-3}$] $= -0.014^{+0.454}_{-0.355}$.
\end{enumerate}
We will first examine how our free parameters are constrained individually be each
observation, and then we will examine the combined constraints.

\subsubsection{\bf The \citet{bouw15} SFR constraints}
\begin{figure}
\setlength{\epsfxsize}{0.45\textwidth}
\centerline{\includegraphics[scale=0.48]{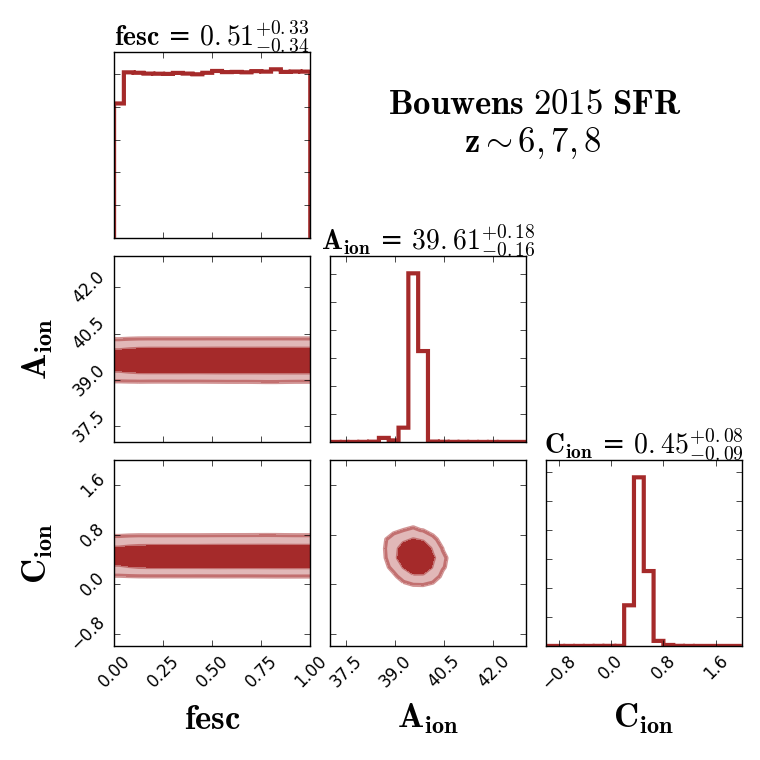}}
\caption{\citet{bouw15} SFR observations, at z=6,7,8 combined together, constraints on our model parameters. Values on top of  the 1-D PDFs diagonal represent the best fit parameters with 1-$\sigma$ (14th and 84th percentiles). Dark and light shaded regions correspond to 1-$\sigma$ and 2-$\sigma$ levels respectively.   The \citet{bouw15} SFR observations provide tight constraints on the $\aion$ and $\cion$ in the selected prior range. As expected, the SFR measurements don't constrain the $\fesc$.}
\label{fig:sfrcons}
\end{figure}

Unlike other semi-numerical models that rely on the efficiency
parameter $\zeta$, our model allows a direct comparison to the SFR
measurements by using a parameterisation for $\rion$ that is directly
relatable to SFR. For a consistent comparison with \citet{bouw15}
measurements, we convert the $\rion$ back to SFR using Equation (2) in \citet{fin11} that is based on ~\citet{sch03} models, and add up all SFR from halos brighter than
$\rm M_{AB} = -17$ at $z=6,7,8$.  To compute the corresponding $\rm M_{AB}$, we use the linear relation provided in \citet{Kenn98} which converts the SFR to luminosity L$_{\nu}$ over the wavelength range 1500-2800 $\rm \AA$. 

Figure~\ref{fig:sfrcons} shows the posterior distribution
of our parameters as constrained solely by  \citet{bouw15} integrated SFR
observations at z=6,7,8 (taken together). This provides somewhat
tigher constraints than fitting to a single redshift of SFR
measurement, although constraining to a single redshift yields
similar results, which indicates that the weak redshift evolution
in the SFR measurements is adequately reproduced by our model for
$\rion$.

The \citet{bouw15} SFR observations provides tight constraints on
the $\aion$ and $\cion$ as seen in figure~\ref{fig:sfrcons}.  while
poorly constraining $\fesc$. The latter is expected because the $\fesc$ is set by the recombinations in 
the ISM while the SFR depends on the halo mass and redshift.

The value of $\cion=0.45$ agrees within the 1-$\sigma$ level with
what was previously found from fitting our hydrodynamical
simulations, which yielded $\cion=0.41$~\citep{hassan16}.  This means that
our large volume semi-numerical model is compatible with the same
slope of the $\rion$-$M_{h}$ relation predicted by the small volume
6/256-RT simulation to match the \citet{bouw15} SFR observation,
thereby nicely corroborating the direct simulation results.

However, the differences are more significant in the $\aion$ posterior
distribution. We see that the $\aion$ best-fit value of 10$^{40.03}$
predicted by 6/256-RT simulation over-estimates by $\sim$ 50\% the
value of $\aion=10^{39.61}$ favoured by our {\sc SimFast21} MCMC
fit using only the \citet{bouw15} SFR constraints.  This represents
somewhat poor concordance at only a 3-$\sigma$ level.  This discrepancy
arises due to the small box size (=6h$^{-1}$Mpc) of 6/256-RT
simulation that does not capture the large scale fluctuations and
massive dark matter halos that contribute significantly to the
reionisation photon budget.  Hence, the 6/256-RT simulation requires
larger $\aion$ to compensate for these limitations. This effect is
also seen in figure~\ref{fig:eorhistory} when comparing the
reionisation histories of the 6/256-RT simulation with our large
volume semi-numerical simulations which tend to reionise the universe
much earlier at a fixed optical depth due to the presence of those
massive halos and large scale-fluctuations.

Overall, utilising only the integrated SFR observations already
gives interesting constraints on the slope and amplitude of the
ionising photon output as a function of halo mass.  However, there
are no useful constraints on the escape fraction.

\subsubsection{\bf The \citet{planck16} optical depth constraints}\label{sec:tau_constraints}

\begin{figure}
\setlength{\epsfxsize}{0.45\textwidth}
\centerline{\includegraphics[scale=0.48]{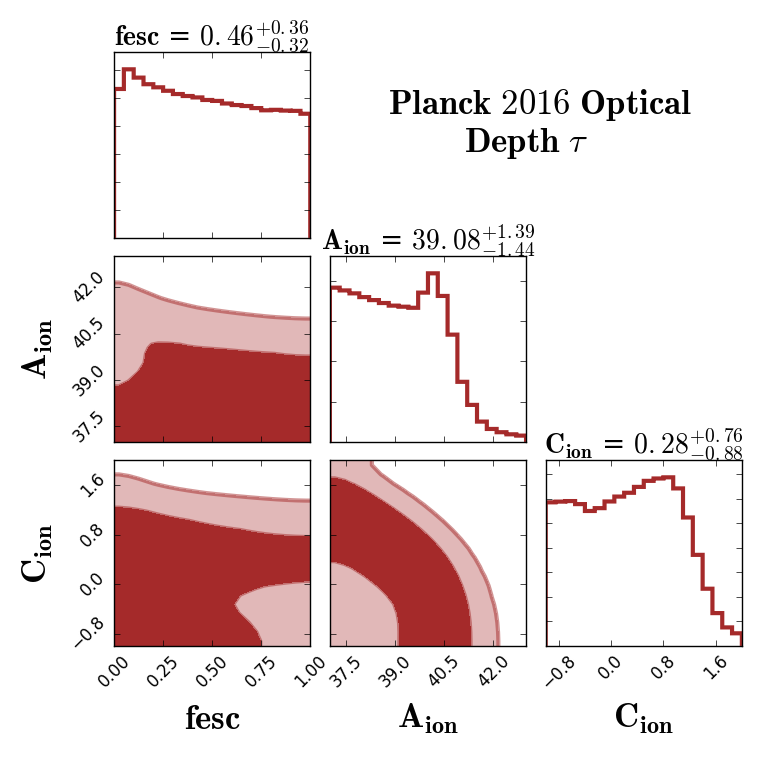}}
\caption{\citet{planck16} optical depth constraints on our model parameters.  Values on top of  the 1-D PDFs diagonal represent the best fit parameters with 1-$\sigma$ (14th and 84th percentiles). Dark and light shaded regions correspond to 1-$\sigma$ and 2-$\sigma$ levels respectively. The \citet{planck16} $\tau$ provides poor constraints on all parameters while there is slight tendency towards lower $\fesc$ values. The \citet{planck16} $\tau$ prefers models with low $\aion$ and $\cion$ values for the chosen prior range as compared to values implied by the \citet{bouw15} SFR observations. }
\label{fig:taucons}
\end{figure}

\begin{figure}
\setlength{\epsfxsize}{0.45\textwidth}
\centerline{\includegraphics[scale=0.45]{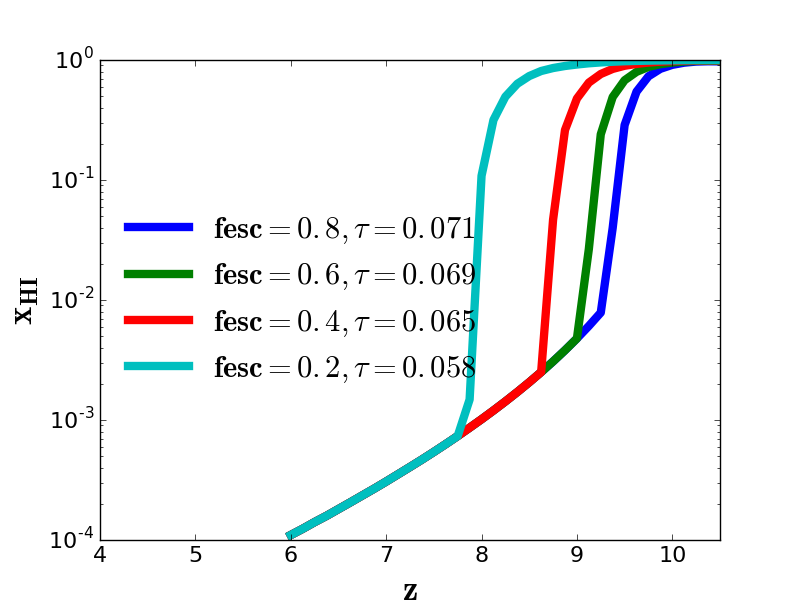}}
\caption{The reionisation history for our Time-integrated model with different $\fesc$ values while fixing the $\aion$ and $\cion$ to values implied by the recent \citet{planck16} $\tau$ measurements. This clearly shows that the current \citet{planck16} optical depth $\tau = 0.058 \pm 0.012$ does not provide tight $\fesc$ constraints for models with rapid reionisation scenarios as the case with our Time-integrated EoR model.}
\label{fig:xHItaucons}
\end{figure}

Figure~\ref{fig:taucons} shows the parameters constrained to match
solely the \citet{planck16} data.  This shows that the Thomson
optical depth data alone provides fairly poor constraints on any
of the parameters.  There is a slight tendency to favour lower
$\fesc$ values, as also found by \citet[see their Figure~4]{Greig16},
but in general all values from zero to one are still allowed.

The main reason for the lack of sensitivity to $\fesc$ is shown in
Figure~\ref{fig:xHItaucons}, and essentially arises from the
still-large errors on $\tau$.   Figure~\ref{fig:xHItaucons} shows
the volume-weighted global neutral fraction evolution for fixed
values of $\aion$ and $\cion$, and shows that $\fesc=20\rightarrow
80$\% gives rise to $\tau=0.058\rightarrow 0.071$, which is still
essentially within the $1\sigma$ uncertainty on the measurement of
$\tau = 0.058 \pm 0.012$.  Hence much smaller error bars on $\tau$
are required to provide better constraints on $\fesc$.

$\aion$ and $\cion$ are also not well constrained by the Thomson
optical depth data alone, though there is some tendency to favour
small values of $\aion$ and $\cion$.  Nonetheless, the uncertainties
are large, and the values favoured from the SFR constraints alone
are within the $1\sigma$ uncertainties of these predictions, as are
the values found directly from the hydrodynamic simulations.

In summary, the Thomson optical depth as measured by {\it Planck}
alone does not provide strong constraints on any of our parameters.
It is clear that reducing uncertainties and/or including other
data will be required in order to meaningfully constrain the sources
driving reionisation.

\subsubsection{\bf The \citet{bec13} ionising emissivity constraints}

\begin{figure}
\setlength{\epsfxsize}{0.45\textwidth}
\centerline{\includegraphics[scale=0.48]{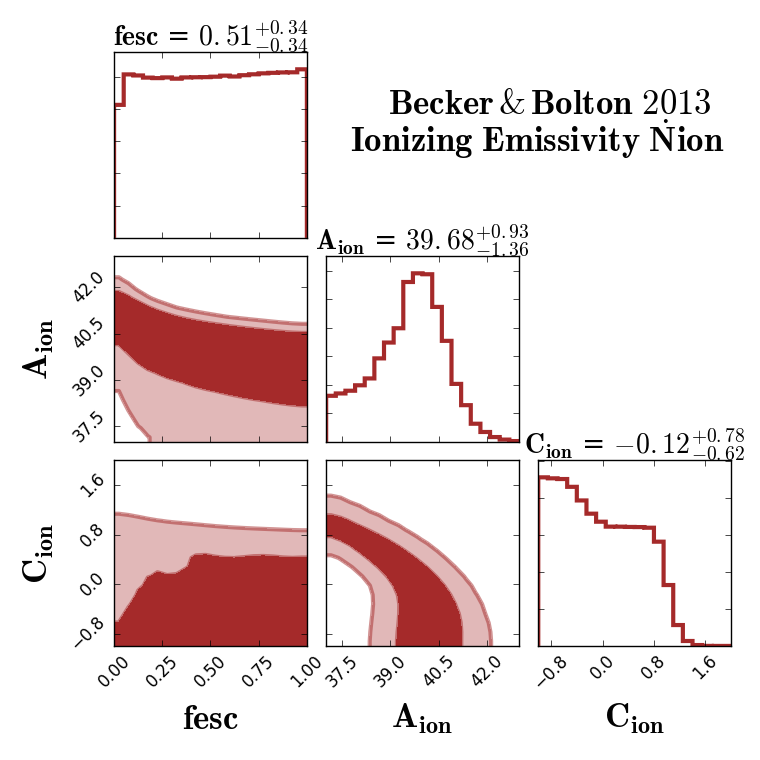}}
\caption{The \citet{bec13} ionising emissivity (z=4.75) constraints on our model parameters. Values on top of  the 1-D PDFs diagonal represent the best fit parameters with 1-$\sigma$ (14th and 84th percentiles). Dark and light shaded regions correspond to 1-$\sigma$ and 2-$\sigma$ levels respectively. Similar to previous constraints, the $\fesc$ is poorly constrained and similar to $\tau$ constraints, the data prefers lower $\fesc$ values. The $\rm \dot{N}ion$  data also prefers models with negative $\cion$ in our selected prior range. This shows that matching to post-reionisation data requires fewer ionising photons and prefers models with dominant contributions from small dark matter halos.}
\label{fig:nioncons}
\end{figure}

The integrated emissivity of ionising photons $\rm \dot{N}_{ion}$
quantifies the total ionisation rate density from all ionising
sources that escape galaxies to fill the intergalactic medium. Mathematically,
$\rm \dot{N}_{ion} = \sum \fesc \rion $ divided by the simulation's comoving volume.   
To compare with \citet{bec13} $\rm \dot{N}ion$ measurements, we add
up $\fesc \rion$ from all halos and divide by the simulation comoving
volume at z=4.75. As with the SFR data, our model permits a direct
comparison with the $\rm \dot{N}ion$ data since we use a parameterisation
for $\rion$ rather than a single efficiency parameter.

Figure~\ref{fig:nioncons} shows the posteriors for our three free
parameters constrained only to match the \citet{bec13} $\rm \dot{N}ion$
data.  As with the SFR and $\tau$ constraints, the $\fesc$ is
unconstrained by this data.  Similar to \citet{planck16} $\tau$
constraints, we find that models with high $\aion$ and $\cion$
values are disfavored by \citet{bec13} $\rm \dot{N}ion$ measurements,
but again this is within $1\sigma$ of the SFR-only constraints.

The slight tendency of $\rm \dot{N}ion$ data towards negative values
of $\cion$ (negative slope of $\rion$-$M_{h}$ relation) favours
small halos being the dominant ionising photon sources contributor
to match the post-reionisation measurements. In contrast, SFR and
$\tau$ data prefers the positive side of $\cion$ values, implying
that massive halos are more important during the reionisation. This
shows that reionisation requires more ionising photons while matching
post-reionisation data requires fewer ionising photons.  This stands
as one of the theoretical challenges for the EoR models as it is not easy to match 
simultaneously observational constraints during and after
reionisations. $\rm \dot{N}ion$ measurements at higher redshifts
($z \sim\, 6,7$) would be very useful to see if this tension extends
into the overlapping redshift regime~\citep{keat14}.

\subsubsection{\bf Combined SFR + $\bf \bf \tau$ + $\bf \bf \rm \dot{N}ion$ constraints}\label{sec:allconst}

\begin{figure}
\setlength{\epsfxsize}{0.45\textwidth}
\centerline{\includegraphics[scale=0.48]{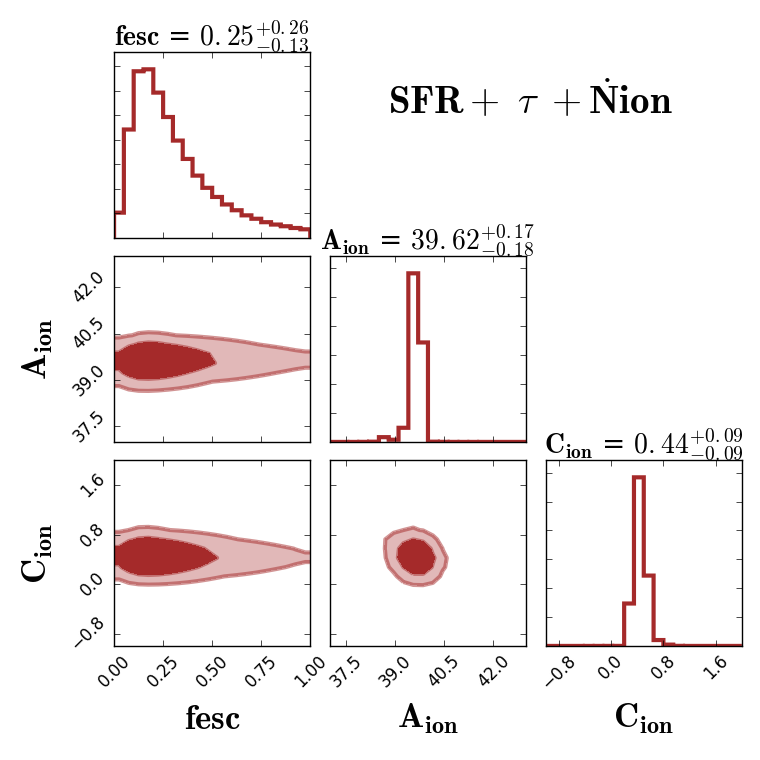}}
\caption{Combined constraints from SFR, $\bf \bf \tau$ and $\bf \bf \rm \dot{N}ion$.  Values on top of  the 1-D PDFs diagonal represent the best fit parameters with 1-$\sigma$ (14th and 84th percentiles). Dark and light shaded regions correspond to 1-$\sigma$ and 2-$\sigma$ levels respectively. Combining all these observations results in a tighter $\fesc$ constraints while $\aion$ and $\cion$ still follow \citet{bouw15} SFR constraints.}
\label{fig:allcons}
\end{figure}

To obtain the strongest constraints given the observations we
consider, we now combine our three key EoR constraints: the
\citet{bouw15} SFR observations,  \citet{planck16} optical depth
measurements and the \citet{bec13} $\rm \dot{N}ion$ data. This
represents the best available constraints we can make given current
data, and serves to provide our base model from which we will do
forecasting for 21cm experiments.

Figure~\ref{fig:allcons} shows the parameter estimates as fit to
the combined sample of these EoR observations. We see that the
$\aion$ and $\cion$ are tightly constrained, which as
Figure~\ref{fig:sfrcons} showed is driven by the \citet{bouw15} SFR
constraints, as the other observations did not provide very
tight constraints on these parameters.

The more interesting difference is in $\fesc$, where the combined
constraints now definitely prefers lower $\fesc$ values, with best
fit-value of $0.25^{+0.26}_{-0.13}$.  This is still a rather wide range,
and the posterior ellipses show that even very low escape fractions
are not ruled out at more than a $\sim 1\sigma$ level, and very
high $\fesc$ values are only disfavoured at $\la 2\sigma$.  This
tendency was hinted at from matching to \citet{bec13} $\rm \dot{N}ion$
and \citet{planck16} optical depth individually.  This result
 indicates that our previous findings of
$\fesc= 0.04-0.06$ in \citet{hassan16} is clearly possible for
models with higher $\aion$ and $\cion$ values within their derived 1-$\sigma$ level. A summary of the
individual and combined constraints is provided in
Table~\ref{constraints_tab}.

This shows that current observations can already constrain the basic
power law parameters of the ionising photon output versus halo mass,
but constraints on $\fesc$ are still somewhat elusive.  Note that
we are also assuming a constant $\fesc$ for all galaxies, while
there may be some mass and/or redshift dependence; however, with
even a single parameter already being poorly constrained, it is
unlikely that adding more parameters will allow tighter constraints.

\begin{figure*}
\setlength{\epsfxsize}{0.5\textwidth}
\centerline{\includegraphics[scale=0.45]{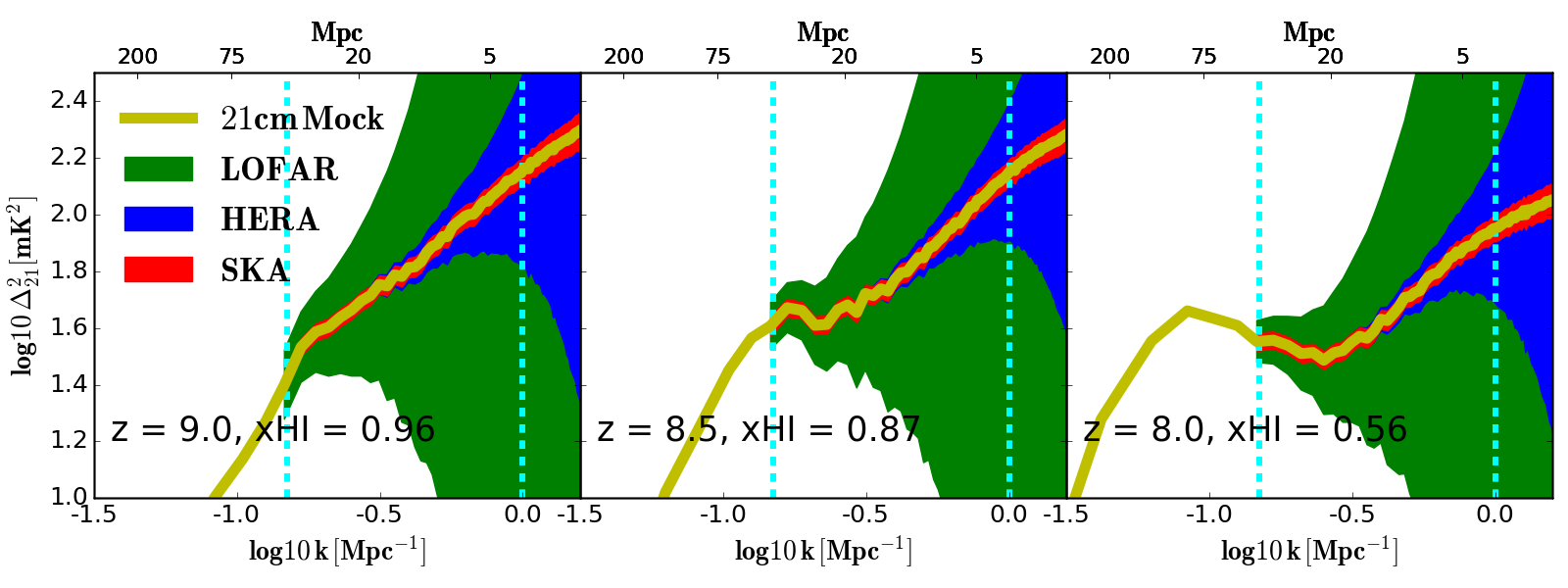}}
\caption{Three redshifts mock 21cm EoR observations using the well-calibrated Time-integrated EoR model with parameters ($\fesc,\,\rm log10(A_{ion}),\, C_{ion}$) = (0.24, 39.63, 0.43). Yellow solid line represents the 21cm power spectrum from the Large box mock observation (L=300/N=560). Shaded area shows the erorrbars obtained using {\sc 21cmSense} package for our constructed EoR arrays: SKA (red), HERA (blue), LOFAR (green). Redshifts and neutral fractions for 21cm mock observations are quoted in each panel. Vertical cyan dashed lines show our chosen k-range to preform the 21cm MCMC.}
\label{fig:21cmmock}
\end{figure*}

\section{21cm forecasting and experiments sensitivities}\label{sec:21cmforecating}

The ultimate goal is to add the 21cm observations to these existing
data (or future improved versions thereof), in order to ascertain
how well we can understand the sources of reionisation.  To do so,
we adopt a forecasting approach by which we use expected uncertainties
from future 21cm power spectrum measurements in concert with these
existing data and ascertain how much improvement the 21cm data
will provide in the precision with which our parameters are
constrained.  We will assume a base model that is the best-fit to
our current constraints as listed in Table~\ref{constraints_tab}.

We focus our analysis on LOFAR, HERA, and SKA1-Low. For each
experiment, we first compute the thermal noise power spectrum which
dominates the errors in measuring the 21cm signal. We then add
more uncertainties from the sample variance, while neglecting the
shot noise since it has been shown to have a minimal effect at the
relevant scales ($k < 2 h$Mpc$^{-1}$) for these telescopes
sensitivities~\citep{pob13}. We obtain these uncertainties using
the {\sc 21cmSense} package\footnote{https://github.com/jpober/21cmSense},
and refer to \citet{par12} for the full mathematical derivation of
the radio interferometer sensitivities, and to ~\citet{pob13,pob14}
for more details on observation strategies and foreground removal
models. We briefly highlight the basic equations and concepts used
in {\sc 21cmSense} to obtain the 21cm power spectrum error from a
specific array configuration.

The dimensionless power spectrum of the thermal
noise~\citep{par12,pob13,pob14} can be obtained using:
\begin{equation}
\Delta^{2}_{N}(k) \approx X^{2}Y \frac{k^{3}}{2\pi^{2}} \frac{\Omega}{2t} T^{2}_{sys}\, ,
\end{equation}
where $X^{2}Y$ is a conversion factor from angle and frequency units
to comoving cosmological distances, $\Omega$ is the primary beam
field-of-view, $t$ is the integration time and $T_{sys}$ is the
system temperature (sky+receiver). It is then straightforward to
add the sample variance to the thermal noise to obtain the total
error~\citep{pob13} as follows:
\begin{equation}
\delta \Delta^{2} (k) = \left(   \sum \frac{1}{ \left( \Delta^{2}_{N}(k) + \Delta^{2}_{21}(k) \right)^{2}}  \right)^{-\frac{1}{2}}\, , 
\end{equation}
where $\Delta^{2}_{21}$ is the 21cm power spectrum and the summation
runs over all measured independent k-modes.

We construct these experiments as follows:

\begin{itemize}

\item {\bf LOFAR}: We use the Netherlands 48 High-Band Antennas
(HBA) with positions listed in~\citet{van13} following~\citet{pob14}.
Each antenna has a diameter of 30.75 m which results in a total
collecting area of 35,762 m$^{2}$ for the 48 HBA station. The
receiver temperature T$_{rcvr}$ is set to 140,000 mK as suggested
by~\citet{jen13,Greig15}.

\item {\bf HERA}: We consider the final design of 331 hexagonally
packed 14 m antennas~\citep{ewall16,bear15}. With this configuration,
the total collecting area becomes 50,953 m$^{2}$. We assume an
100,000 mK receiver temperature T$_{rcvr}$ , similar to previous
works by ~\citet{pob14,Greig15}.

\item {\bf SKA-LOW1}: We model SKA1-Low following the SKA1 System
Baseline Design document by \citet{dew} in which the proposed array
consists of 911 antennae in total. These antennae are distributed
randomly to form a compact core using 866 dishes surrounded by the
remaining 45 dishes along spiral arms. The 866 core antennae 
provide the vast majority of the sensitivity, and hence our
SKA model ignores those 45 spiral arms stations~\citep{pob14,Greig15}. 
Each station of 866 antennae has a
diameter of 35 m which makes a total collecting area of 833,189
m$^{2}$. The receiver noise here is determined by: T$_{rcvr}$ = 0.1
T$_{sky}$ + 40 K, where the sky temperature is modelled using:
T$_{sky}$  = 60$\lambda^{2.55}$.  
\end{itemize} 

For a consistent comparison, we choose to operate these three array
designs in a drift-scanning mode for 6 observing hours per day
for 180 days at 8 MHz bandwidth.  We consider
\citet{pob14} moderate foreground removal model where the foreground
wedge extends 0.1 h Mpc$^{-1}$ beyond horizon limit.

\begin{table*} \LARGE
 \scalebox{0.75}{\begin{tabular}{ || l || c || c || c ||}\hline
	& ${\rm\bf  \fesc}$ & ${\rm\bf  log10(A_{ion})}$ & ${ \rm\bf C_{ion} }$\\ \hline \hline
	   {\bf 21cm Mock Observations } & & &\\ \hline \hline
    {\bf SKA} & 0.240$^{+0.056}_{- 0.054}$  & 39.628$^{+0.030}_{- 0.032}$  & 0.431$^{+0.052}_{- 0.056}$ \\ \hline 
   {\bf  HERA} & 0.237$^{+0.061}_{-  0.054}$ & 39.626$^{+0.031}_{-0.025}$  & 0.425$^{+0.055}_{- §0.058}$ \\ \hline 
	 {\bf LOFAR} & 0.415$^{+0.384}_{- 0.239}$  & 39.229$^{+0.606}_{- 1.117}$  & 0.445$^{+0.341}_{-  0.274}$ \\ \hline \hline
	 {\bf 21cm Mock Observations + ALL (SFR, $\rm\bf  \dot{N}_{ion}$, $\bf\rm \tau$)}& & & \\ \hline \hline   
    {\bf  SKA+ALL} & 0.217$^{+0.052}_{- 0.048}$  & 39.631$^{+0.024}_{- 0.029}$  & 0.423$^{+0.053}_{- 0.057}$ \\ \hline   
    {\bf  HERA+ALL} & 0.221$^{+0.058}_{- 0.051}$  & 39.630$^{+0.029}_{- 0.029}$  & 0.427$^{+0.053}_{- 0.059}$ \\ \hline   
    {\bf  LOFAr+ALL} & 0.206$^{+0.069}_{- 0.045}$  & 39.634$^{+0.042}_{- 0.033}$  & 0.421$^{+0.065}_{- 0.060}$ \\ \hline   
   
\end{tabular}}
\caption{Summary of our parameter estimations from the 21cm mock observations and from combining the 21cm mock observations with the current EoR observations (SFR, $\rm\bf  \dot{N}_{ion}$, $\bf\rm \tau$). }\label{constraints_tab2}
\end{table*}

\subsection{Including the 21cm data}

We combine three different redshifts of 21cm power spectrum
observations, namely $z= 9.0, 8.5, 8.0$, which provides tighter
constraints than considering any single epoch observations.  With
multiple redshifts 21cm observations, one accounts simultaneously
for the variation in redshift (density field) and neutral fraction
(ionisation field) evolution, which are the main components in
determining the 21cm fluctuations. Given the rapid reionisation
behaviour of the Time-integrated model as shown in figure~\ref{fig:eorhistory},
our selected redshifts ($z=9,8.5,8$) correspond to a wide range of
neutral fractions that account for different reionisation epochs
such as the initial bubble growth and the bubble overlap phase.  We
next construct the likelihood from these observations by simply
adding up their individual $\chi^{2}$. We limit our analysis to a
wide $k$-range of 0.15-1.0 Mpc$^{-1}$, consistent with \citet{Greig15}.
From this $k$-range, we select 10 bins of the power spectrum which
is sufficient to capture the fluctuations for a given 21cm power
spectrum.

We use the well-calibrated Time-integrated model with parameters derived
from fitting to our combined set of EoR observations as discussed
in \S~\ref{sec:allconst} and shown in figure~\ref{fig:allcons}. Specifically, we
use the following parameters: ($\fesc,\,\rm log10(A_{ion}),\,
C_{ion}$) = (0.24, 39.63, 0.43), consistent with the 1-$\sigma$
level of constraints by our combined set of EoR observations. We
then use these parameters to create our mock observations with a
large box size of $L=300$ Mpc and $N=560$ per side which results in a
resolution of 0.375 $h^{-1}$ Mpc. We determine the error in measuring
the 21cm power spectra for our mock observation by using the
telescope sensitivity code {\sc 21cmSense} for each specific array
experiments at our chosen redshifts as described above. We use the
same pipeline discussed in \S~\ref{sec:pipeline} to sample the 21cm
power spectrum space, except now we include the 21cm mock
observation power spectra among the pre-computed runs to study how
well the MCMC technique may recover the input model parameters.

Figure~\ref{fig:21cmmock} shows our 21cm mock observations at several
redshifts. The shaded area corresponds to the error in measuring
the 21cm power spectrum for our large mock observations using the
{\sc 21cmSense} package. LOFAR (green shading), operating currently,
will be able to constrain only the largest scales considered here,
while HERA (blue), under construction now, will be further sensitive
to intermediate scales, while the future SKA1-Low (red) will provide
tight constraints into the sub-Mpc scale regime owing to its wider
baselines and hence better resolution.  These uncertainties depend
mainly on our telescope configurations as described above.  Hence
the main improvement as these facilities develop will be to better
constrain the 21cm power spectrum towards smaller scales, and each
generation will provide significant gains in this.

\begin{figure}
\setlength{\epsfxsize}{0.5\textwidth}
\centerline{\includegraphics[scale=0.45]{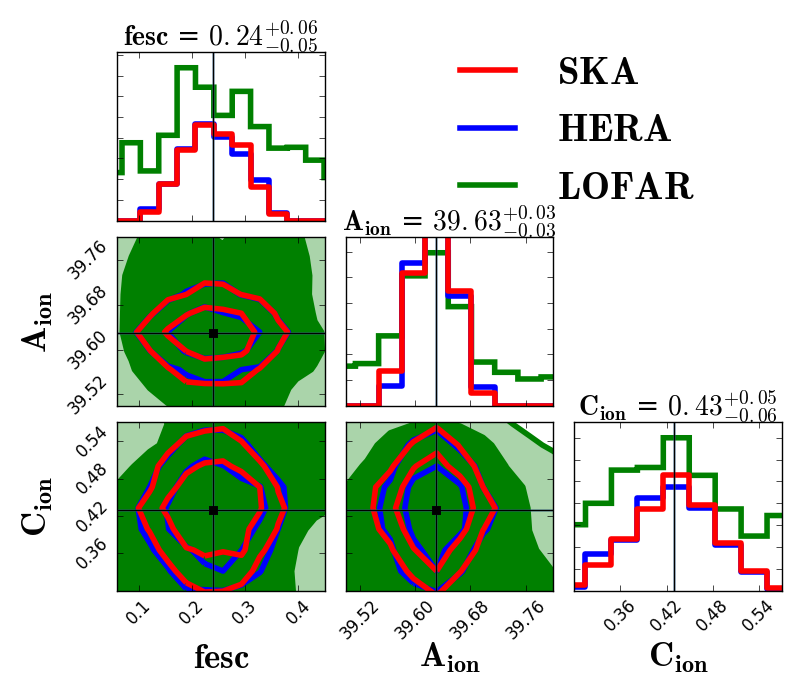}}
\caption{21cm power spectrum constraints on our three EoR parameters from several redshifts (z=9.0,8.5,8.0) mock observations. SKA, HERA, and LOFAR constraints are shown by red, blue, and green contours respectively. Values on top of the 1D PDFs represent the best fit parameters as implied by the SKA mock observations while black square points correspond to the input mock observation parameters: ($\fesc,\,\rm log10(A_{ion}),\, C_{ion}$) = (0.24, 39.63, 0.43). The MCMC technique is able to recover the input model parameters. It is evident that the future 21cm observations can tightly constrain our model parameters for experiments with small and intermediate levels of uncertainty in detecting the expected signal such as SKA and HERA respectively. }
\label{fig:21cmnoall}
\end{figure}

\begin{figure}
\setlength{\epsfxsize}{0.5\textwidth}
\centerline{\includegraphics[scale=0.44]{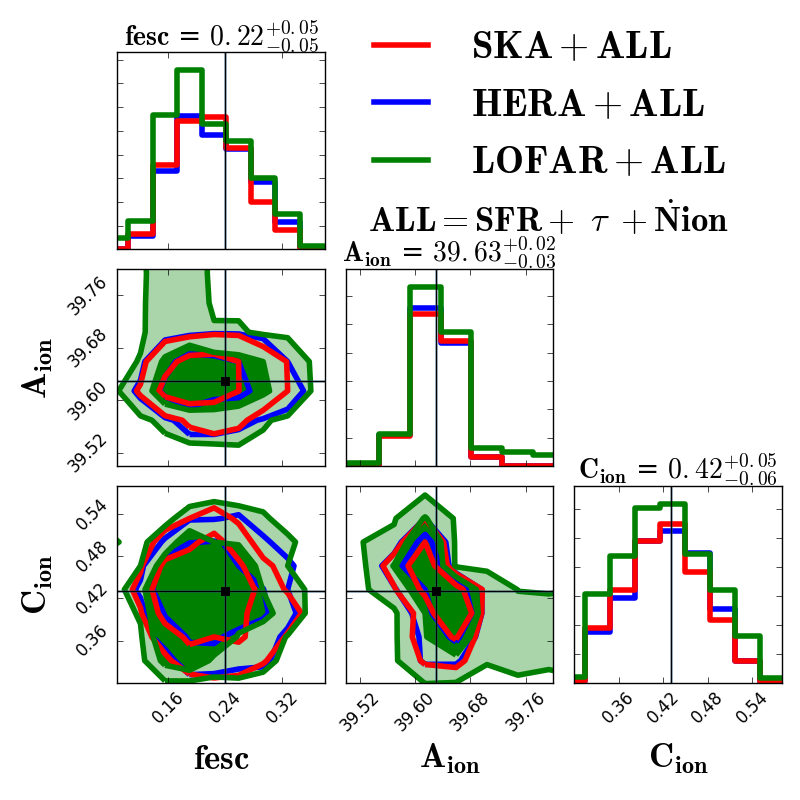}}
\caption{Parameter estimates from combining the current EoR observations with the 21cm mock observations. The current EoR observations here are our previous combined sample:  SFR, $\bf \bf \tau$ and $\bf \bf \rm \dot{N}ion$ while the 21cm mock observations are combinations of several 21cm redshifts at z=9.0,8.5,8.0. SKA, HERA, and LOFAR constraints are shown by red, blue, and green contours respectively. Values on top of the 1D PDFs represent the best fit parameters as implied by the SKA mock observations while black square points correspond to the input mock observation parameters: ($\fesc,\,\rm log10(A_{ion}),\, C_{ion}$) = (0.24, 39.63, 0.43). It is evident that adding the current EoR observations on top of the 21cm mock observations provide more tighter constraints even for experiments with large 21cm power uncertainties such as the case with LOFAR.}
\label{fig:21cmwall}
\end{figure}

\subsection{21cm MCMC}\label{sec:21cmmcmc}

We now ask how well the 21cm data can constrain our free parameters.
First, we consider the 21cm power spectrum data as shown in
Figure~\ref{fig:21cmnoall} by itself, to see how tightly our
parameters can be constrained by such observations alone.  Then we
add the 21cm data to our other existing observational constraints.
In each case we use our MCMC framework to determine our best-fit
values of our free parameters and their uncertainties using the
entire data set, for the case of each telescope facility.  This
provides forecasting for how much improvement can be expected from
future 21cm observations.

Figure~\ref{fig:21cmnoall} shows the 1D PDFs and 2D contours of our
three parameters from the combined redshifts ($z=9.0,8.5,8.0$) of
21cm mock observations by our three selected EoR experiments. To
begin, we see that our MCMC search well recovers the best-fit input
model (mock observation) parameters (black square points).
This is to be expected, since this same input model was
used to generate the 21cm data. The improvement to be noted here
is the reduction of the uncertainties on these parameters relative
to the previous case without 21cm data.

For LOFAR (green shaded area), we see that the 21cm observations don't
provide tight constraints due to large uncertainties as
seen in figure~\ref{fig:21cmmock}.  Essentially, mildly constraining
the large-scale power provides little information on the ionisation
sources that drive reionisation.

In contrast, HERA (blue) and SKA (red) provide quite tight constraints
on the free parameters.  Note that the scale of the posteriors is
substantially reduced relative to our previous plots in order to
enhance visibility.  Hence future 21cm data alone can already
independently constrain reionising sources, without adding in any
other observations.  Interestingly, there is almost no difference
between the SKA and HERA constraints.  This arises because the
parameter constraints are predominantly driven by the larger scales,
and HERA and SKA provide similar constraints on the power spectrum
for scales $\ga 2$~Mpc.

Comparing the 21cm constraints with constraints obtained from
combining several EoR key observables, we find that constraining
to 21cm observations yield smaller parameter errors. This can
be clearly seen when comparing the 1-$\sigma$ level of $\fesc$ and
$\aion$ found by constraining to the 21cm observations (fig~\ref{fig:21cmnoall}) versus to
the combined EoR sample (SFR, $\rm \dot{N}_{ion}$, $\tau$) (fig~\ref{fig:allcons}). However,
it is evident that the 21cm future observation can constrain the $\fesc$
tighter than the current EoR key observables.

In previous work by ~\citet{Greig15}, the authors used a similar
semi-numerical framework and performed similar analysis to constrain
their free parameters to future 21cm mock observations. However,
they did not have the photon escape fraction as a free parameter
and rather constrained their efficiency parameter $\zeta$, from
which the $\fesc$ can be computed for various assumptions about gas
fraction in stars and ionising photons number per baryons (see their
eq. (2)). However, we here constrain the $\fesc$ directly without
making further assumptions about the gas and baryons fractions,
hence our presented $\fesc$ results are direct, albeit the inherent
photon conservation issues in these semi-numerical models, which
we will discuss later.

We finally constrain our free parameters by combining the 21cm mock
observations with the current EoR key observables (SFR, $\rm
\dot{N}_{ion}$, $\tau$) as shown in figure~\ref{fig:21cmwall}. From
this figure, we see that our three parameters are
well-constrained by the combined set of current EoR and 21cm mock
observations. Adding our combined EoR sample (SFR, $\rm \dot{N}_{ion}$,
$\tau$) on top of the 21cm mock observations improves the error in
estimating our free parameters, particularly for arrays with large
21cm errorbars such as LOFAR. This shows that the future 21cm
observations are important in constraining the model astrophysical
parameters and complement the other existing EoR various observations.
A summary of our  21cm mock observations constraints combined
with the other EoR observations is given in table~\ref{constraints_tab2}.

Our 21cm forecasting shows that the future 21cm power spectrum
observations will be crucial for providing tight constraints on
various parameters related to the sources of reionisation.  Even
by themselves, such data will provide improve constraints over what
can be obtained using current observations.  When combined with
other observations, the constraints get quite tight, even for the
difficult-to-constrain photon escape fractions $\fesc$.  The tightness
of the constraints suggest that it may be possible to independently
constrain variations in the escape fraction with mass or redshift;
we will examine this in future work.

\subsection{Photon conservation}

To make use of our $\fesc$ constraints, we here test the photon
conservation problem in our semi-numerical model. Previous
semi-numerical models, based on the excursion set formalism, have
pointed out a violation in the photon number conservation. In
\citet{zah07}, the authors found that their semi-numerical model
loses about 20\% photons. They have argued that this photon loss
arises from ionised bubbles overlapping, which they compensated by
boosting the efficiency parameter $\zeta$. More recent work by
\citet{paran16} have developed a Monte Carlo Model of bubble growth
to resolve the photon conservation problem in their semi-numerical
model. Although their bubble growth model didn't resolve the problem
completely, nevertheless improvements have been achieved and
they have demonstrated that the problem comes from the fact that
the excursion set-based models use the average mass of the bubbles
rather than tracking the actual mass of sources and bubble local
density fluctuations.

\begin{figure}
\setlength{\epsfxsize}{0.5\textwidth}
\centerline{\includegraphics[scale=0.48]{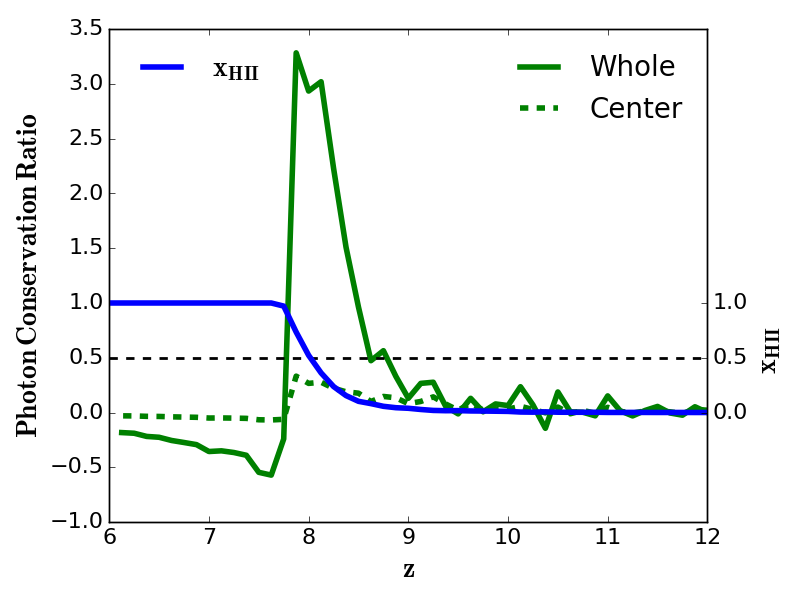}}
\caption{Photon conservation ratio from the Time-integrated EoR model using whole flagging (green solid) versus center flagging  (greed dashed) scheme, with their reionisation history (blue solid). Horizontal black dashed line represents the 50\% neutral/ionised fraction limit. Both methods violate the photons number conservations as seen by the under-ionisation at high redshifts and over-ionisation at end of reionisation.  }
\label{fig:pho_con}
\end{figure}

However, there are two methods to flag the spherical regions as ionised in the excursion set-formalism. The first is to flag the whole cells in the bubble (whole flagging) whereas the second is to flag only the center cell of the bubble (center flagging). We next use these two methods to verify the photon conservation in our Time-integrated EoR model. We would expect that, during time interval $\rm dt$, the total number of escaped ionising photons ($\rm \fesc\rion\,dt$) minus the total number of recombinations ($\rm \rrec\,dt$)  should be equal to the number of ionisations in the neutral hydrogen atoms ($\rm (x_{HII}(t_{i+1}) - x_{HII}(t_{i}) )  N_{H}$). In other words, the successful photons that manage to escape from the interstellar medium (corrected by $\fesc$) and from high density regions along the way (subtracted by $\rm \rrec\,dt$) should be equal to the total number of neutral atoms that have been ionised during time $\rm dt$. We can write the photon conservation ratio as follows:
\begin{equation}
\rm Photon\, Conservation\, Ratio\,  =  \frac{(x_{HII}(t_{i+1}) - x_{HII}(t_{i}) )  N_{H}}{ (\fesc\rion - \rrec )\,dt},
\end{equation}
where $\rm dt = t_{i+1}-t_{i}$. This ratio should be equal to unity for an ideal photon conserving model. However, the ratio can be less than unity when the universe is highly ionised.  We then apply this ratio to the two methods, whole flagging versus center flagging, to check the photon conservation problem in both. We note that center flagging scheme requires about 20\% more ionising photons to match the reionisation history obtained by  whole flagging method. We then adjust the $\fesc$ in two methods to reproduce identical reionisation history (identical $\tau$) while keeping other parameters fixed. 

In figure~\ref{fig:pho_con}. we plot the photon conservation ratio for the two methods, whole flagging  (green solid line) and  center flagging (green dashed line) with the reionisation history (blue solid line). We find that the  center flagging scheme under-uses photons during all reionisation redshifts, even after reionisation (z $<$ 8), which might partly explain the need for higher $\fesc$ with this method. The photon loss in the whole flagging scheme agrees qualitatively with center flagging at higher redshifts when the universe is almost neutral. 

As reionisation proceeds, the whole flagging starts to over-use photons and ionises more neutral atoms than expected. The photon excess/loss in the two methods are clearly redshift dependent. In the center flagging method, the photon loss is by a factor of $\sim$ 3,7,20 at z= 7.75, 9.25, 11 respectively. The whole flagging scheme shows photon loss (under-using photons) at high redshifts and photon excess (over-using photons) at the end of reionisation. At high redshifts, the photon loss, in the whole flagging, is by a factor of $\sim$ 2,4,7 at z= 8.75,9,11 respectively. This shows that, at high redshifts, the photon loss, in the whole flagging method, is less by a factor of $\sim$ 2,3 as compared with center flagging method. At z=8.5 ($\rm x_{HI} \sim$ 0.9), the whole flagging method satisfies the photon conservation condition as the ratio becomes unity, but the ratio does not converge at unity afterwards. After this point, the whole flagging scheme starts to overuse photons increasingly by a large amount till the end of reionisation. We find the photon excess is about 10\% at z=8.4 and 70\% at z=7.75 (end of reionisation).

We note that all our EoR models adopts the whole flagging method. This shows that our constrained photon escape fractions $\fesc$ are, in fact, over-estimated by the photon excess associated with the whole flagging method. All our previous $\fesc$ estimations can be corrected and lowered by 10\% up to 70\% depending on redshifts. The photon loss/excess evolution in redshift suggests that the $\fesc$ might be required to change with redshift in order to preserve photon number conservation as a temporary solution.

\section{conclusion}\label{sim:conclusion}

We have improved our {\sc SimFast21} semi-numerical code for computing
the EoR on large scales by incorporating a more physically-motivated
criterion for determining whether a region of space is ionised, as
well as integrating our framework into a full MCMC parameter search
framework so we can forecast how well current and future observations
can constrain the physical properties of the sources driving
reionisation.

We have calibrated our new model to various current observations
of the EoR, namely the \citet{bouw15} SFR observations, the
\citet{planck16} optical depth measurements, and the \citet{bec13}
$\rm \dot{N}ion$ data.  We also compared our new EoR model to our
previous EoR model in \citet{hassan16} in terms of their EoR history,
\ion{H}{ii} bubble sizes, and 21cm power spectra.  We further studied
variations in the 21cm fluctuations produced by all possible variants
of our ionisation conditions.

We then presented a robust MCMC analysis to constrain our generalized
source model's free parameters against current EoR observations.
We used the well-calibrated EoR model to predict the 21cm power
spectrum for the future EoR array experiments SKA, HERA, and LOFAR.
We show how the future 21cm observations are important for complementing
the existing EoR current observations in order to tightly estimate
the astrophysical parameters of EoR sources.

Our key findings are as follows:

\begin{itemize}

\item The Time-integrated EoR model produces very large \ion{H}{ii} bubbles
as compared with the Instantaneous EoR model, and fewer small bubbles.
This difference is clearly shown in their evolving HI maps
(figure~\ref{fig:lightcones}) and the ionisation field (figure~\ref{fig:ion_maps}). This results in a larger ionisation and 21cm power spectrum on large scales by 1-1.2 orders of magnitude as seen in (Figure~\ref{fig:pkxi} and ~\ref{fig:pk21}).

\item By considering all possible combinations between the hydrogen
atoms and recombination terms in the ionisation condition, we showed
that recombinations are subdominant in determining the 21cm power
spectrum particularly on large scales (Figure~\ref{fig:pk21models}).
The 21cm power spectrum amplitude and shape are highly sensitive
to accounting for the amount of and fluctuations in the neutral
hydrogen density. This means semi-numerical models must carefully
account for the neutral hydrogen to robustly predict the expected
21cm signal.

\item The \citet{bouw15} SFR observations provide tight constraints
on the ionising emissivity amplitude $\aion$ and the slope of the
$\rion$-M$_{h}$ relation $\cion$, but provide no constraint on the
photon escape fraction $\fesc$ (Figure~\ref{fig:sfrcons}). The recent
 \citet{planck16} optical depth (Figure~\ref{fig:taucons}) and the \citet{bec13} $\rm
\dot{N}ion$ measurements (Figure~\ref{fig:nioncons}) poorly constrain our model parameters,
while they slightly prefers models with lower values of $\fesc$, $\aion$ and $\cion$.

\item Combining all of SFR, $\tau$ and $\rm \dot{N}ion$ together
results in tighter parameter constraints, as seen in
Figure~\ref{fig:allcons}. The $\aion$ and $\cion$ here follow the
previous constraints by the SFR observations, but combining these
measurements yields better escape fraction constraints of
$\fesc=0.25^{+0.26}_{- 0.13}$, though still not very tight. The
parameters determined directly from the full hydrodynamic simulations
analysed in \citet{hassan16} are consistent with these constraints.

\item Using the well-calibrated Time-integrated EoR model, we predict the
21cm power spectrum at different redshifts ($z=9,8.5,8$) for several
constructed radio array designs, namely SKA, HERA, and LOFAR
(Figure~\ref{fig:21cmmock}).  While LOFAR does not provide strong
constraints except at the largest scales, future experiments will
tightly constrain the 21cm power spectrum to smaller scales that
can better constrain the reionising source population.

\item By adding current EoR observations (SFR, $\tau$, $\rm
\dot{N}ion$) to the 21cm mock observations, we find that all
experiments recovers the input model parameter accurately and the
parameters error are futher improved.  This illustrates how future
21cm observations can complement and substantially improve upon
existing EoR observations in order to more tightly constrain the
emissivity of EoR sources and their relationship to the underlying
halo population.

\item We find that photon conservation is sub-optimal owing to the
way the excursion set formalism is generically implemented in current
semi-numerical codes, including {\sc SimFast21}.  The root difficulty
is that cells are treated as fully neutral or fully ionised, with
no possibility of intermediate ionisation levels.  While some tuning
could be done to minimise the problem, a robust solution likely
lies in replacing the excursion set-formalism with a proper
photon-conserving radiative transfer approach.  We leave this for
future work.
\end{itemize}
We have discussed the uncertainty associated with
ionization condition in the excursion set-based models and found that a 
slight change in the ionization condition  could lead to a big difference 
in the 21cm power spectrum particularly on large scales as seen between the 
Time-integrated and Instantaneous model. A possible approach to break such 
degeneracy and resolve the ionization condition uncertainty is to compare 
these models' 21cm power spectra to radiative transfer simulations. For this reason,
we are currently developing our own radiative transfer routine ({\sc SimFast21-RT}) and the result will be 
forthcoming.

The {\sc SimFast21-MCMC} platform developed here will be applicable
for a wide range of EoR forecasting science cases.  By robustly
incorporating all the current observables within an MCMC framework
and being able to straightforwardly incorporate new data, we are
building the tools necessary to optimally connect future redshifted
21cm power spectrum from the EoR to physical quantities associated
with the population of reionising sources.  Such a framework can
be used to explore and constrain exotic source populations such as
mini-quasars or Population~III stars, as well as to potentially
extend to multi-tracer cross-correlation approaches.  The most
immediate hurdle will be to develop a more robust yet still fast
radiative transfer method that conserves photons, so we can more
reliably assess the obtained source population parameters.  There
is much exciting work to be done as we continue to prepare for the
21cm EoR era.

 \section*{acknowledgements}
We thank Jonathan Pober for making his {21cmSense} sensitivity code
publicly available, providing the SKA antennas coordinates, helpful discussions and comments. 
The authors acknowledge helpful discussions with Andrei Mesinger,
Greig Bradley, Jonathan Zwart, Girish Kulkarni, Tirthankar Roy Choudhury, and Neal Katz.
We also thank Dan Foreman-Mackey for his excellent MCMC code {\sc EMCEE} and 
visualisation package {\sc corner} \citep{foreman16}. 
SH is supported by the Deutscher Akademischer Austauschdienst (DAAD) Foundation.  RD
and SH are supported by the South African Research Chairs Initiative
and the South African National Research Foundation. MGS is supported by the South
African Square Kilometre Array Project and National Research Foundation. This work was also supported
by NASA grant NNX12AH86G.  Part of this work was conducted at the
Aspen Center for Physics, which is supported by National Science
Foundation grant PHY-1066293.  Computations were performed at the
cluster ``Baltasar-Sete-Sois'', supported by the DyBHo-256667 ERC
Starting Grant, and the University of the Western Cape's ``Pumbaa"
cluster.

\appendix
\section{REIONIZATION HISTORY CONVERGENCE TEST}\label{sec:contests}

\begin{figure}
\setlength{\epsfxsize}{0.5\textwidth}
\centerline{\includegraphics[scale=0.48]{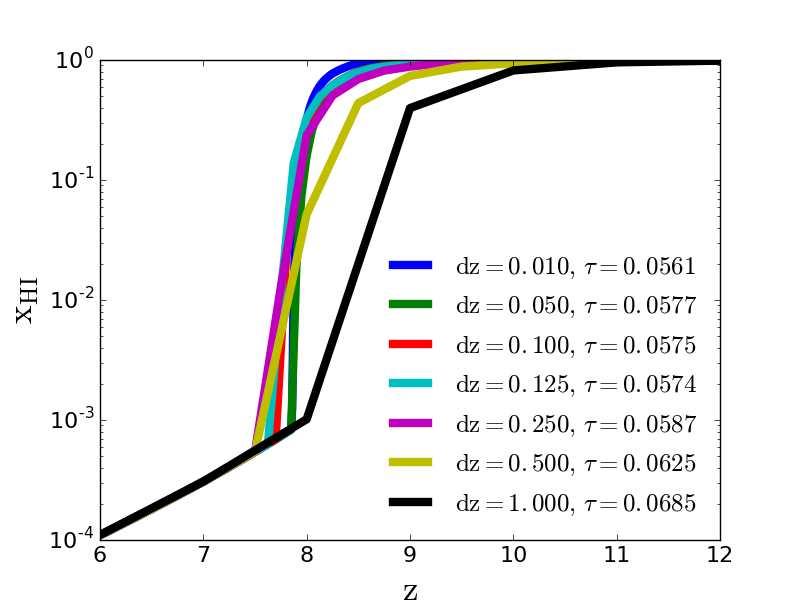}}
\caption{Reionisation history convergence test in our new Time-integrated EoR model using different time interval steps as quoted in the legend along with the corresponding optical depth. It is evident that our model is well converged for dz $\leqslant$ 0.125.  }
\label{fig:converge_xHI}
\end{figure}

As mentioned earlier, our new Time-integrated EoR model is sensitive
to the choice of simulation time step dz, due to implementing
time-integrated ionisation condition (equation~\ref{eq:new_con}).
The correct reionisation history is however achieved for very small
values of dz (ideally when dz goes to zero), which is computationally
impossible. We here present a convergence test to support our choice
of dz=0.125. Figure~\ref{fig:converge_xHI} shows the well-calibrated Time-integrated
model reionisation history for different time steps dz using a
simulation box size of L=75 Mpc and N=140$^{3}$. It is clear that
the reionisation starts earlier (higher $\tau$) when adopting larger
steps (dz = 1.0 in black or 0.5 in yellow) than using smaller step
values (dz $\leqslant$ 0.25). This means using larger dz permits
more ionising photons production, and hence lower neutral fractions
as seen in the beginning of the reionisation.

Comparing the reionisation history obtained with dz = 0.125 (cyan)
versus that with dz = 0.01 (blue, the most correct result), we find
both models produce similar tau ($\rm \Delta \tau \sim 0.001$) and
the reionisation history shape is identical. There is a very minor
difference in the neutral fraction of $\rm \Delta x_{HI} \sim 0.01$
by end of reionisation. Hence we conclude that our Time-integrated
model is numerically well converged for dz $\leqslant$ 0.125.

\end{document}